\documentclass{emulateapj}

\usepackage{amssymb,amsmath,mathrsfs}
\usepackage{latexsym}
\usepackage{fancyhdr}
\usepackage{graphicx}
\usepackage{inputenc}
\usepackage{fontenc}
\usepackage{textcomp}
\usepackage{color}

\usepackage{natbib}
\usepackage{amsmath}
\usepackage{epsfig}

\shorttitle{Pulsation Frequencies and Modes of Giant Exoplanets}
\shortauthors{Bastien Le Bihan \& Adam Burrows}

\begin{document}

\title{Pulsation Frequencies and Modes of Giant Exoplanets}
\author{Bastien Le Bihan\altaffilmark{1} \& Adam Burrows}
\affil{Department of Astrophysical Science, Peyton Hall
Princeton University, Princeton, NJ 08544, USA.}
\altaffiltext{1}{Ecole Polytechnique,
Palaiseau, France.}
\email{bastien.le-bihan@polytechnique.edu, burrows@astro.princeton.edu}
%\tableofcontents

\begin{abstract}

We calculate the eigenfrequencies and eigenfunctions of the acoustic oscillations of giant exoplanets and explore the dependence of the characteristic frequency $\nu_0$ and the eigenfrequencies on several parameters: the planet mass, the planet radius, the core mass, and the heavy element mass fraction in the envelope. We provide the eigenvalues for degree $l$ up to 8 and radial order $n$ up to 12. For the selected values of $l$ and $n$, we find that the pulsation eigenfrequencies depend strongly on the planet mass and radius, especially at high frequency. We quantify this dependence through the calculation of the characteristic frequency $\nu_0$ which gives us an estimate of the scale of the eigenvalue spectrum at high frequency. For the mass range $0.5 \leq M_P \leq 15$ $M_J$,  and fixing the planet radius to the Jovian value, we find that $\nu_0 \sim 164.0 \times \left(M_P/M_J\right)^{0.48} \mu Hz$, where $M_P$ is the planet mass and $M_J$ is Jupiter's mass. For the radius range from 0.9 to 2.0 $R_J$, and fixing the planet's mass to the Jovian value, we find that $\nu_0 \sim 164.0 \times \left(R_P/R_J\right)^{-2.09} \mu Hz$, where $R_P$ is the planet radius and $R_J$ is Jupiter's radius. We explore the influence of the presence of a dense core on the pulsation frequencies and on the characteristic frequency of giant exoplanets. We find that the presence of heavy elements in the envelope affects the eigenvalue distribution in ways similar to the presence of a dense core. Additionally, we apply our formalism to Jupiter and Saturn and find results consistent with both the observationnal data of \citet{gaulme11} and previous theoretical work.

\end{abstract}

\section{Introduction}

Pulsation frequencies and modes are potentially useful tools with which to study the interior structure of the giant planets. Three types of modes are distinguished: g-modes are standing internal gravity waves, p-modes are standing acoustic waves, and f-modes are of intermediate frequency and can be regarded as the fundamental mode of either the p- or the g-modes. The corresponding frequencies are characterized by their radial order $n$ and degree $l$. 

The surface movements associated with such pulsations are hard to detect and, as of July 2012, only Jupiter's global velocity oscillations have been observed. The work of \citet{schmider91}, \citet{mosser91}, \citet{mosser93}, and \citet{mosser00} resulted in a putative measurement of the mean frequency spacing of 142 $\pm$ 3 $\mu Hz$ \citep{mosser00}. More recently, \citet{gaulme11} claims to have detected Jupiter’s global modes with a mean noise level five times lower than previously achieved, using observations acquired in 2005 by the SYMPA Fourier spectro-imager of the Teide Laboratory. Their upper troposphere radial velocities are determined by measurements of the Doppler shifts of solar Mg lines (517 nm) reflected by Jupiter's clouds. The resulting velocity maps were decomposed into spherical harmonics to create a set of time series whose power was computed with a discrete Fourier transform. It exhibited excess power between 800 and 2000 µHz and a secondary excess between 2400 and 3400 µHz, with a frequency of maximum amplitude of 1213 $\pm$ 50 $\mu Hz$, a mean spacing of 155.3 $\pm$ 2.2 $\mu Hz$ and a mode maximum amplitude of $49^{+8}_{-10} \ cm\ s^{-1}$. These measurements agree with theoretical expectations in terms of the frequency range, the amplitude, and the mean large spacing \citep{bercovici87, provost93} and correspond to the signature of p-modes.

Several theoretical works intended to bring out possible forcing mechanisms capable of exciting pulsation modes in nearby gaseous planets to magnitudes accessible to observation. Current studies tend to struggle to find significant theoretical amplitudes, but the crucial point is that giant exoplanets may exhibit more favorable conditions for these mechanisms,  which may bring about higher oscillation amplitudes. Indeed, \citet{bercovici87} and \citet{marley90, marley91} evaluated the possibility of a coupling of acoustic oscillations to turbulent convection on Jupiter and Saturn, as is the case for the Sun \citep{goldreich94}. With this kind of coupling, the physical amplitudes of the modes may scale like $L M^{\alpha}$, where $L$ is the interior luminosity, $M$ is the Mach number, and $\alpha$ is a power that depends on whether the sound generated is via dipole ($\alpha = 3$) or quadrupole ($\alpha = 5$) emission \citep{bercovici87}. Giant exoplanets will be almost fully convective, and their internal luminosities are likely to be 10 to 100 times larger than that of Jupiter \citep{burrows01}. Since the Mach number is also likely to be higher, the amplitudes of acoustic oscillations of giant exoplanets may exceed those of Jupiter by a large factor. Moreover, close encounters of planets in an evolving planetary system may promote planet-planet or moon-planet interactions which can excite tides dynamically through the transfer of orbital energy, as pointed out in the stellar case by \citet{lee86} and in the moon case by \citet{marley90}. If the encounters are close enough, then the resultant amplitudes may be significant.

Given current technological limitations to the observation of giant planet's oscillations, it is reasonable to think that the global oscillations of giant exoplanets will not be detected in the foreseeable future. However, specific environnments of giant planets outside the solar system may foster excitation mechanisms more vigorous than in the Jovian or Saturnian cases and, thus, lead to significant pulsation amplitudes. Hence, a theoretical exploration of the systematics of the pulsation frequencies for the broad spectrum of recently discovered giant exoplanets might stimulate observers to design methods to detect giant planet oscillations, since such oscillations are so diagnostic of structure.

In this spirit, we calculate the eigenfrequencies and eigenfunctions of the pulsationnal modes of giant exoplanets. We quantify the dependence of the modal eigenfrequencies on the planet mass and radius. In addition, we focus on the influence of a dense core on these quantities, since its presence has already been suggested in specific extrasolar giant planets \citep{burrows07, guillot06}. Furthermore, we calculate corresponding models for Jupiter and Saturn themselves, and compare them with previous work.

\citet{vorontsov76}, \citet{vorontsovzarkhov81a, vorontsovzarkhov81b} and \citet{vorontsov81} added rotation, differential rotation, and ellipticity to their initial spherically-symmetric, nonrotating Jovian models. The influence of the troposphere on the high-frequency oscillations was first adressed by \citet{vorontsov89}, then in detail by \citet{mosser94}. \citet{provost93} developed an asymptotic method to determine the eigenfrequencies which included the discontinuity of a Jovian core. They introduced the mean spacing or characteristic frequency $\nu_0$, defined by: $\nu_0 = \left[2 \int_{planet} \frac{dr}{c_0}\right]^{-1}$, where $c_0$ is the speed of sound, and emphasized the sensitivity of the Jovian oscillation spectrum to the presence of a dense core. Since then, the Jovian characteristic frequency has been estimated to be between 152 and 160 $\mu Hz$ \citep{provost93, gudkova95, gudkova99}. These estimates are consistent with the recent observations of \citet{gaulme11}.

Saturn's oscillations have also been studied theoretically. \citet{marley90} suggested that the f-modes of Saturn are the most likely to be detected through their potential influence on that planet's rings. Using the techniques applied to Jupiter, \citet{gudkova95} calculated the eigenfrequencies of the lowest-order p-modes of Saturn, along with their characteristic frequencies. The latter were found to be between 106  and 109 $\mu Hz$.

Throughout our paper, we do not take into account the effects of rotation or oblateness \citep{vorontsovzarkhov81a, vorontsovzarkhov81b, lee93}. Since the adiabatic approximation is appropriate for Jovian planets \citep{marley90}, adiabaticity is here assumed for Jupiter, Saturn, and the entire set of exoplanets. Finally, since we focus on the giant exoplanet regime, we select the appropriate range of measured giant planet radii \citep{udry07}: $0.8 \ R_J \lesssim R_P \lesssim 2.1 \ R_J$.

\citet{gaulme11} and \citet{schmider91} suggest that the lowest-degree p-modes are the most likely to be detected.
\citet{gaulme11} and \citet{marley90} take the degree $l=8$ to be an upper limit and
the Jovian observations of \citet{gaulme11} are within the frequency regions [0.8, 2.1] $mHz$ and [2.4, 3.4] $mHz$.
In the specific case of Jupiter, and for $l \in [0,8]$, these values loosely correspond to the ranges of radial order $n \in [0,12]$ and [15,20], respectively.
These observationnal windows are consistent with the theoretical value of the atmospheric cutoff frequency of Jovian modes estimade by \citet{mosser95} and which is about 3 mHz.
Theoretically, the asymptotic trends are manifest for $n \geq 4$ or 5 \citep{provost93, marley91}. 
Thus, we focus on the p-modes and f-modes of low degree ($l \leq 8$) and relatively low radial order ($n \leq 12$).

In \S 2, we summarize the theory of adiabatic nonradial oscillations of nonrotating spherical planetary models. We present our numerical technique, closely based on the work of \citet{unno89} and \citet{dalsgaard97}.
To test the validity and precision of our code, we calculate the eigenfrequencies of f-modes, p-modes and g-modes of well-studied polytrope models and compare our results to those of \citet{dalsgaard94}.

In \S 3, we describe the giant planet models used in the article. We present our results in the case of Jupiter and Saturn, and compare them to both observationnal data \citep{gaulme11} and theoretical work \citep{provost93, mosser94, gudkova99}.
We briefly focus on the dependence of the Jovian modal oscillations on the core mass of Jupiter and present the derivatives of the low-degree acoustic modes eigenfrequencies with respect to the core mass.

In \S 4, we focus on the giant exoplanets. We present our results in terms of the characteristic frequency $\nu_0$,  the eigenfrequencies of low-degree f-modes, and the eigenfrequencies of low-degree p-modes across the giant exoplanet continuum. In separate subsections, we investigate their dependence on the planet mass, planet radius, and core mass. We also focus on the influence of a high fraction of heavy elements in the envelope, by using a high helium mass fraction, $Y$, as an approximate substitute \citep{spiegel11}. Finally, we briefly discuss the temporal evolution of the charateristic frequency $\nu_0$ for simple planetary models considered in isolation.

%%%%%%%%%%%%%%%%%%%%%%%%%%%%%%%%%%%%%%%%%%%%%%%%%%%%%%%%%%%%%%%%%%%%%%%%%%%%%%%%%%%%%%%%%%%%%%%%%%%%%%%%%%%%%%%%%%%%%%%%
%%%%%%%%%%%%%%%%%%%%%%%%%%%%%%%%%%%%%%%%%%%%%%%%%%%%%%%%%%%%%%%%%%%%%%%%%%%%%%%%%%%%%%%%%%%%%%%%%%%%%%%%%%%%%%%%%%%%%%%%

\section{Methodology and Techniques}

\subsection{Nonradial Oscillation Eigenvalue Problem}

We considered non-rotating, spherically-symmetrical planetary models.
For adiabatic, nonradial oscillations of such objects, it follows from \citet[chap. 13]{unno89} that the radial part of the displacement  $\xi_r$,  the Eulerian perturbations of the pressure $p'$, and the Eulerian perturbation of the gravitational potential $\Phi'$ take the form:
\begin{equation}
f(t,r,\theta,\phi) = f(r) Y^m_l(\theta,\phi)e^{i\sigma t},
\end{equation}
where $Y^m_l$ are the spherical harmonics of azimuthal order $m$ and degree $l$, and $f$ is either $\xi_r$, $p'$, or $\Phi'$. A given oscillation mode is, thus, described by its azimuthal order $m$, degree $l$, radial order $n$.

These variables are govern by a set of differential equations and four boundary conditions, two at the surface, and two at the center \citep{unno89}.
The corresponding set of equations is given in the first section of the Appendix.
Since the azimuthal order $m$ does not appear in the governing equations, the eigenfrequencies are (2$l$+1)-fold degenerate, and are fully described by their degree $l$ and radial order $n$.

This problem has to be numerically implemented to calculate the corresponding eigenfrequency $\nu_{n,l}$ for a given mode. We detail the technique used in this paper in the Appendix, but summarize our overall methodology in the next subsections.

%%%%%%%%%%%%%%%%%%%%%%%%%%%%%%%%%%%%%%%%%%%%%%%%%%%%%%%%%%%%%%%%%%%%%%%%%%%%%%%%%%%%%%%%%%%%%%%%%%%%%%%%%%%%%%%%%%%%%%%%

\subsection{Numerical Implementation}

Several numerical techniques have been previously introduced \citep{vorontsov76, unno89, dalsgaard97}. They are divided into shooting techniques and relaxation methods.
In the shooting technique, solutions satisfying the boundary conditions are integrated separately from the inner and outer boundaries, and the eigenvalue is found by matching these solutions at an arbitrary interior fitting point.
The second technique is to solve the equations together with the normalization condition, and all but one of the boundary conditions, using a relaxation technique;
the eigenvalue is then found by requiring that the remaining boundary condition be satisfied.
The shooting methods are generally considerably faster than the relaxation techniques, but their precision decreases as the degree $l$ increases \citep{dalsgaard97, dalsgaard03}. 
However, since we consider only low-degree modes, a shooting method is quite suitable for our problem.
Dimensionless variables are introduced (see the second section of the Appendix). In particular, the dimensionless frequency $\omega$ is defined by:
\begin{equation}
\omega^2 = \frac{\sigma^2 R^3}{GM} = \frac{4 \pi^2\nu^2 R^3}{GM}. \label{DQ4}
\end{equation}
where $\sigma$ is the angular frequency of the modes, $\nu$ is the corresponding frequency, $G$ is the gravitational constant, and $R$ and $M$ are the radius and the mass of the studied object, respectively.
Solutions are obtained by integration using a fifth-order Runge-Kutta technique.
To calculate the eigenfrequencies in a given frequency range, we use a determinant method developed  in \citet{dalsgaard97} and fully described in the third section of the Appendix.
Two linearly independent solutions are calculated from the center, and two from the surface. They are connected at an arbitrary inner boundary. The eigenvalues do not depend on its position.

									%%%%%%%%%%%%%%%%%%%%%%%%%%%%%%%%%%%%%%%%%%%%%%%%%%%%%%%%%%%%%%%%%%%%%%

\subsection{Mode Order}

As we calculate the eigenfrequencies and eigenfunctions of the acoustic modes for a given degree $l$, we determine their order $n$  using the following equation \citep{dalsgaard97}:
\begin{equation}
n = \sum_{x_{z1} > 0} sign\left(y_2 \frac{dy_1}{dx}\right) + n_0 \ , \label{order}
\end{equation}
where $y_1$ and $y_2$ are dimensionless variables defined by:
\begin{eqnarray}
y_1 & = & \frac{\xi_r}{r},  \nonumber \\
y_2 & = & \frac{1}{gr} \left(\frac{p'}{\rho} + \Phi'\right) ,  \nonumber \\
\end{eqnarray}
where $\rho$ is the density and $g$ is the gravitational acceleration.

In the definition of $n$, the sum is over the zeros ${x_{z1}}$ in $y_1$ (excluding the center), where $x=r/R$ is the relative radius. The value of $n_0$ depends on the behavior of the solution close to the innermost boundary.
If $y_1$ and $y_2$ have the same sign at the innermost
mesh point, excluding the center, $n_0 = 0$. Otherwise $n_0 = 1$. In particular, for a complete model
that includes the center, as in our case, it follows from the boundary conditions at the center that $n_0 = 1$ for
radial oscillations and $n_0 = 0$ for non-radial oscillations.
With these conventions, the order of the f-mode is $n = 0$.

									%%%%%%%%%%%%%%%%%%%%%%%%%%%%%%%%%%%%%%%%%%%%%%%%%%%%%%%%%%%%%%%%%%%%%%

\subsection{Results for a Polytrope Model}

In order to test the code, we compute the eigenfrequencies and eigenfunctions of polytropic models.
To compare our results with the work of \citet{dalsgaard94}, we take the same radius and mass for our calculations: $R_P  =  6.9599 \times 10^{10} \ cm$ and $M_P = 1.989 \times 10^{33} \ g$.
The eigenfrequencies of f-modes, g-modes, and p-modes are given in Tables \ref{pmodes1}, \ref{pmodes2} and \ref{gmodes}, in $\mu Hz$.
Our frequencies match those of \citet{dalsgaard94} to a precision of $10^{-5}$ or better, for all types of modes, for $l \in [0,3]$ and $n \in [-20,25]$.
This gives us confidence in our calculational method as we approach more complex models.

%%%%%%%%%%%%%%%%%%%%%%%%%%%%%%%%%%%%%%%%%%%%%%%%%%%%%%%%%%%%%%%%%%%%%%%%%%%%%%%%%%%%%%%%%%%%%%%%%%%%%%%%%%%%%%%%%%%%%%%%
%%%%%%%%%%%%%%%%%%%%%%%%%%%%%%%%%%%%%%%%%%%%%%%%%%%%%%%%%%%%%%%%%%%%%%%%%%%%%%%%%%%%%%%%%%%%%%%%%%%%%%%%%%%%%%%%%%%%%%%%

\section{Jupiter and Saturn}

\subsection{The Models}

The giant planet models we use for the calculations for Jupiter and Saturn consist of an adiabatic atmosphere, a hydrogen-helium envelope, and an olivine core. When a core is included,  we explore $0 \leq M_{core} \leq 10$ $M_{\oplus}$ 
for Jupiter and $9 \leq M_{core} \leq 22$ $M_{\oplus}$ for Saturn.
Both ranges are marginally consistent with the core accretion formation models for these planets, which suggest $10-20M_{\oplus}$ \citep{saumon04, pollack96}.
The hydrogen/helium equation of state that we use for this study is described in \citet{saumon95}.
The transition between the atmosphere and the envelope has been smoothed to ensure the continuity of density, pressure, and sound speed.

We build several models for both Jupiter and Saturn, using different core masses and helium fractions in the envelope. 
Table \ref{parameters} presents various parameters of these models: the helium mass fraction inside the envelope, $Y$, 
the mass of the core, $M_{core}$ (in Earth units), the central pressure $p_c$ (in Mbars), the central density $\rho_c$ (in cgs units), and the characteristic frequency $\nu_0$ (in $\mu Hz$), defined in \citet{provost93} using the following equation:
\begin{equation}
\nu_0 = \left[2 \int_{planet} \frac{dr}{c_0}\right]^{-1} \ , \label{nu0}
\end{equation}
where $c_0 =  \sqrt{(dp_0/d\rho_0)_{ad}}$ is the sound speed in hydrostatic equilibrium.

\begin{figure}
\includegraphics[width = 0.5\textwidth]{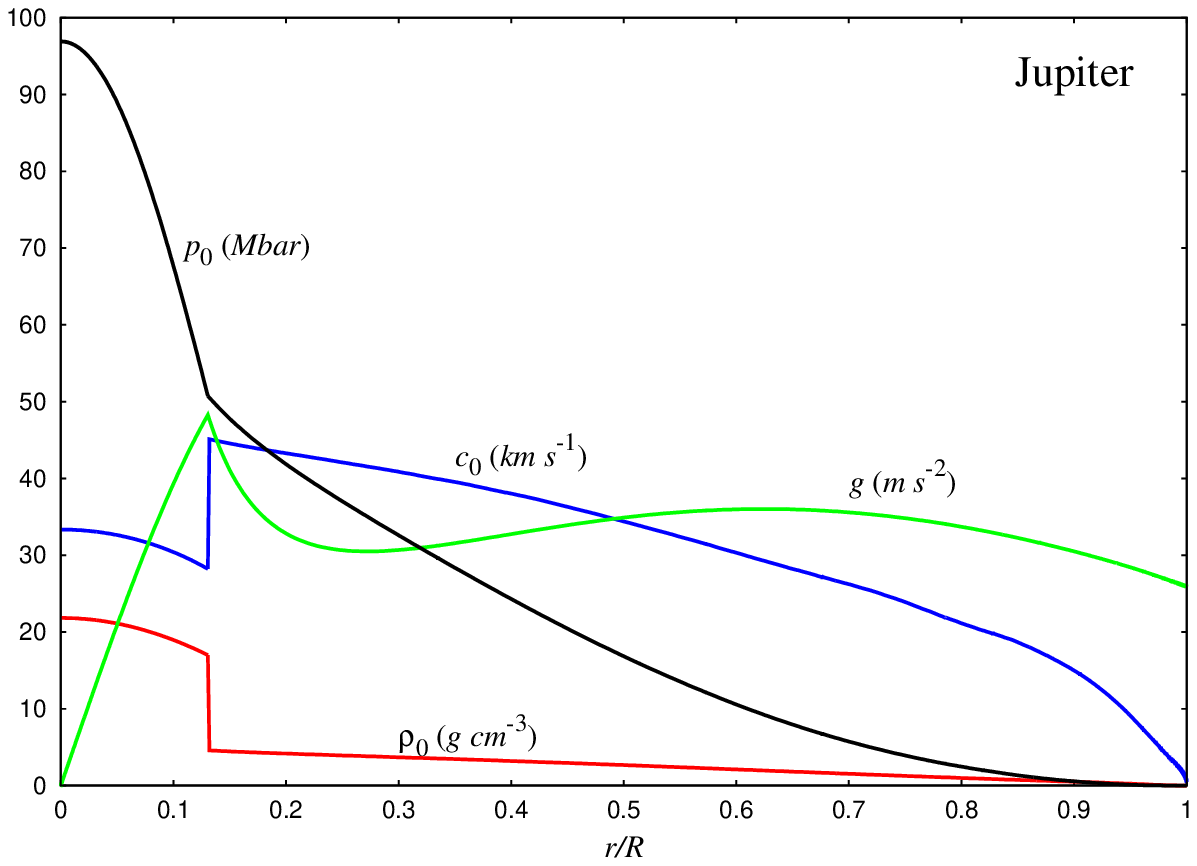}
\includegraphics[width = 0.5\textwidth]{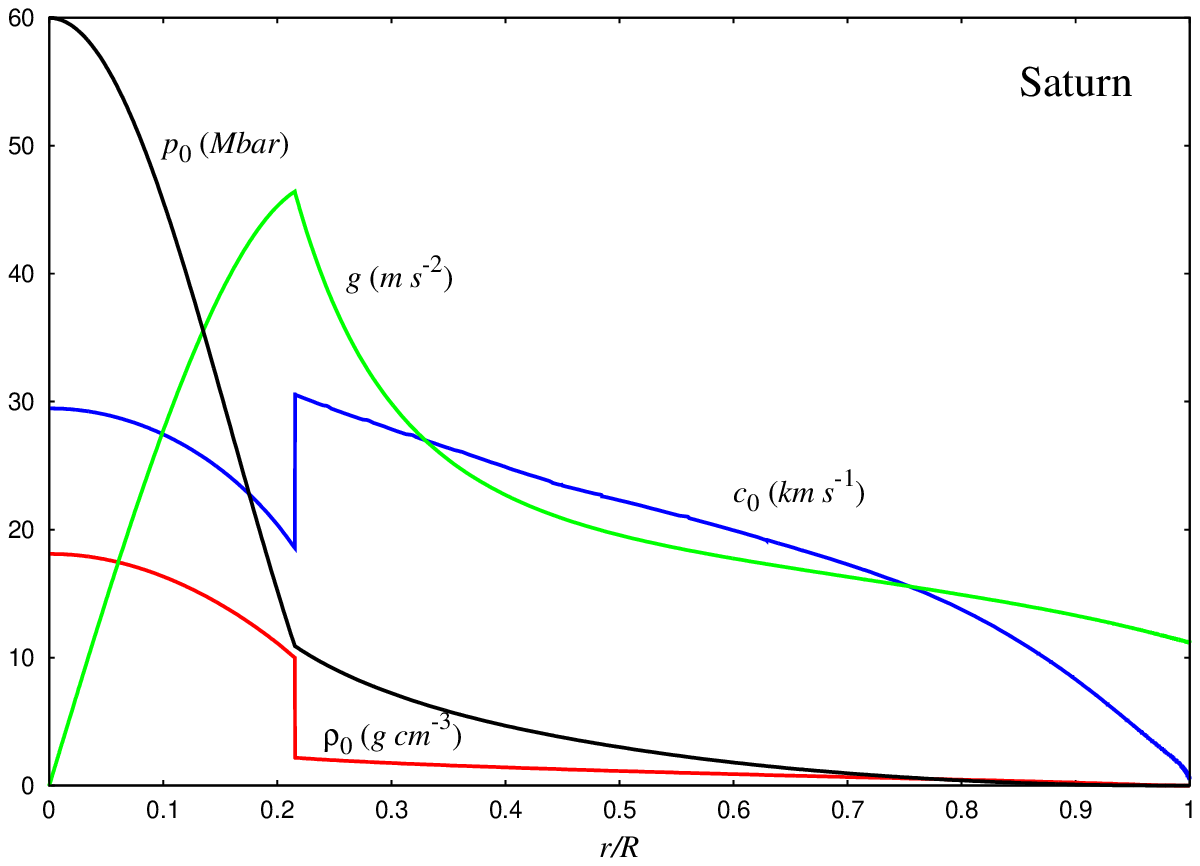}
\caption{Left panel: Distribution of density $\rho_0$ $(g \ cm^{-3})$, gravitational acceleration $g$
$(cm \ s^{-2})$, pressure $p_0$ $(Mbar)$, and  sound speed $c_0$ $(km \ s^{-1})$ as a function of the
relative radius $r/R$ for the model J4 of Jupiter.
Right panel: The same in the case of the model S2 of Saturn.
One has to notice that the scales of the Y-axis are different on the two figures. 
\label{f1}
}
\end{figure}

Figure  \ref{f1} portrays the profiles of the density, $\rho$, the gravitational acceleration, $g$, and the sound speed, $c_0$, for models J4 and S2, defined in Table \ref{parameters}. Both density and sound speed are discontinuous at the core interface.
In the case of Jupiter, for the models defined in Table \ref{parameters}, the calculated characteristic frequencies, $\nu_0$, are consistent with the observational value measured by \citet{gaulme11}: $\nu_0 = 155.3$ $\pm 2.2$ $\mu Hz$ (see also Figure \ref{f6}).

%%%%%%%%%%%%%%%%%%%%%%%%%%%%%%%%%%%%%%%%%%%%%%%%%%%%%%%%%%%%%%%%%%%%%%%%%%%%%%%%%%%%%%%%%%%%%%%%%%%%%%%%%%%%%%%%%%%%%%%%

\subsection{Oscillation Modes \label{JupModes}}

\begin{figure}
\includegraphics[width = 0.5\textwidth]{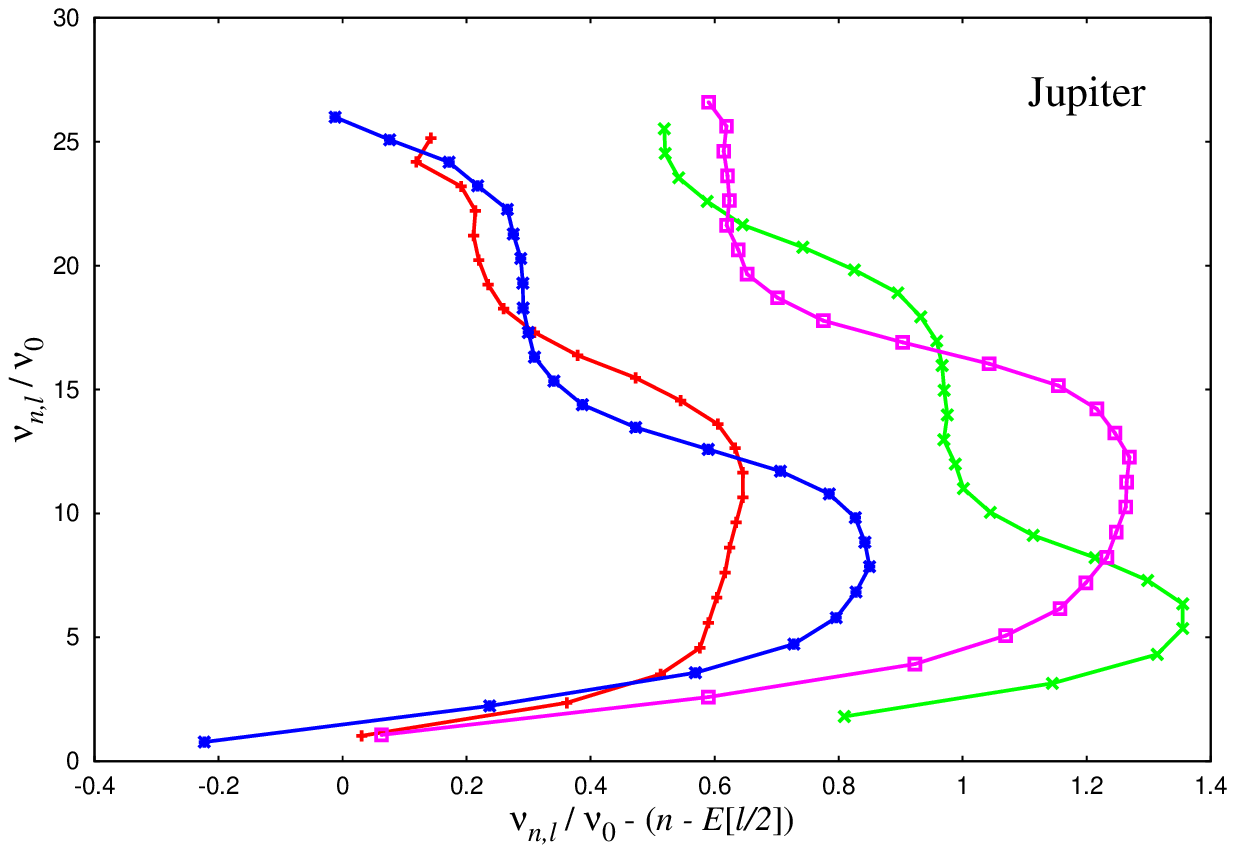}
\includegraphics[width = 0.5\textwidth]{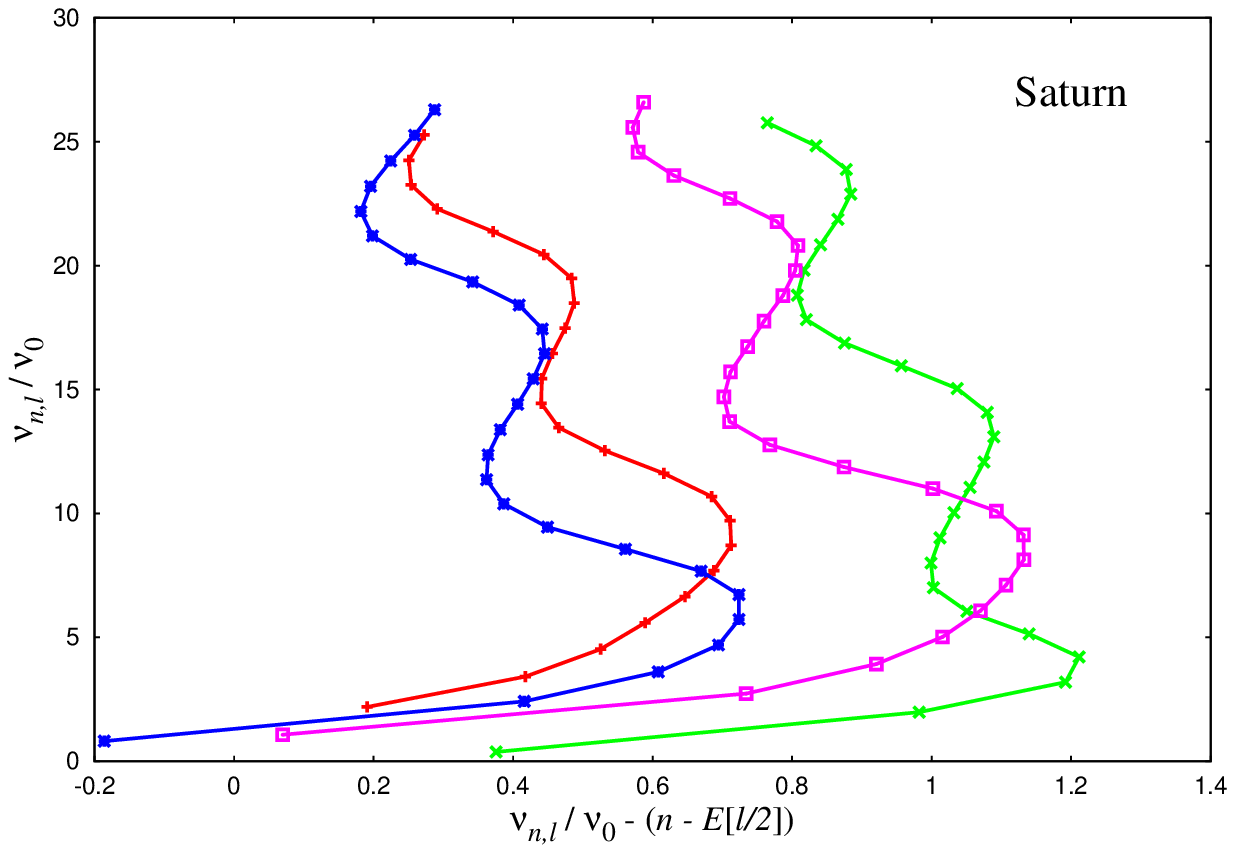}
\caption{Left panel: Echelle diagrams of the eigenfrequencies of Jupiter (model J4) for $l \in [0,3]$ and $n \in [0,25]$. The characteristic frequency, $\nu_0$, is $151.2$ $\mu Hz$  ($l = 0$: $+$,  $l = 1$: $\times$, $l = 2$: $\ast$, $l = 3$: $\boxdot$).
Right panel: The same in the case of the model S2 of Saturn. The characteristic frequency, $\nu_0$, is $115.0$ $\mu Hz$.
\label{f2}
}
\end{figure}

Figure \ref{f2} depicts the eigenfrequencies of models S2 and J4 for $l \in [0,3]$ and $n$ up to 25. These results are given in the form of \'echelle diagrams based on the results of the asymptotic theory for low-degree oscillations developed in, for example, \citet{provost93}. This theory predicts that, for low degree $l$ and large radial order $n$, the eigenfrequencies $\nu_{n,l}$ of p-modes are, to a first approximation, proportionnal to the characteristic frequency, $\nu_0$:
\begin{equation}
\nu_{n,l} \simeq \left(n+ \frac{l}{2} \right) \nu_0 \ .
\end{equation}
An \'echelle diagram presents the ratio $\nu_{n,l}/\nu_0$ as a function of the difference $\nu_{n,l}/\nu_0 - (n + E[l/2])$, where $E$ is the floor function. Thus, it allows us to see the deviation from the approximate value.
Both panels of Figure \ref{f2} are in qualitative agreement with previous numerical results for Jupiter \citep[and their Fig. 3d, Fig. 5a, \& Fig. 4 respectively]{provost93, mosser94, jackiewicz12} and for Saturn \citep[their Fig. 5b]{mosser94}. The Jovian periods of the acoustic fundamental tone and overtones with radial order and degree up to 5 are given in Table \ref{pmodesJupiter} for the model J4. Though the differences between our models and those derived by \citet{gudkova99} prevent us from a precise comparison, it is clear that we find similar results to this previous work (see their Table 2).

\begin{figure}
\includegraphics[width = 0.5\textwidth]{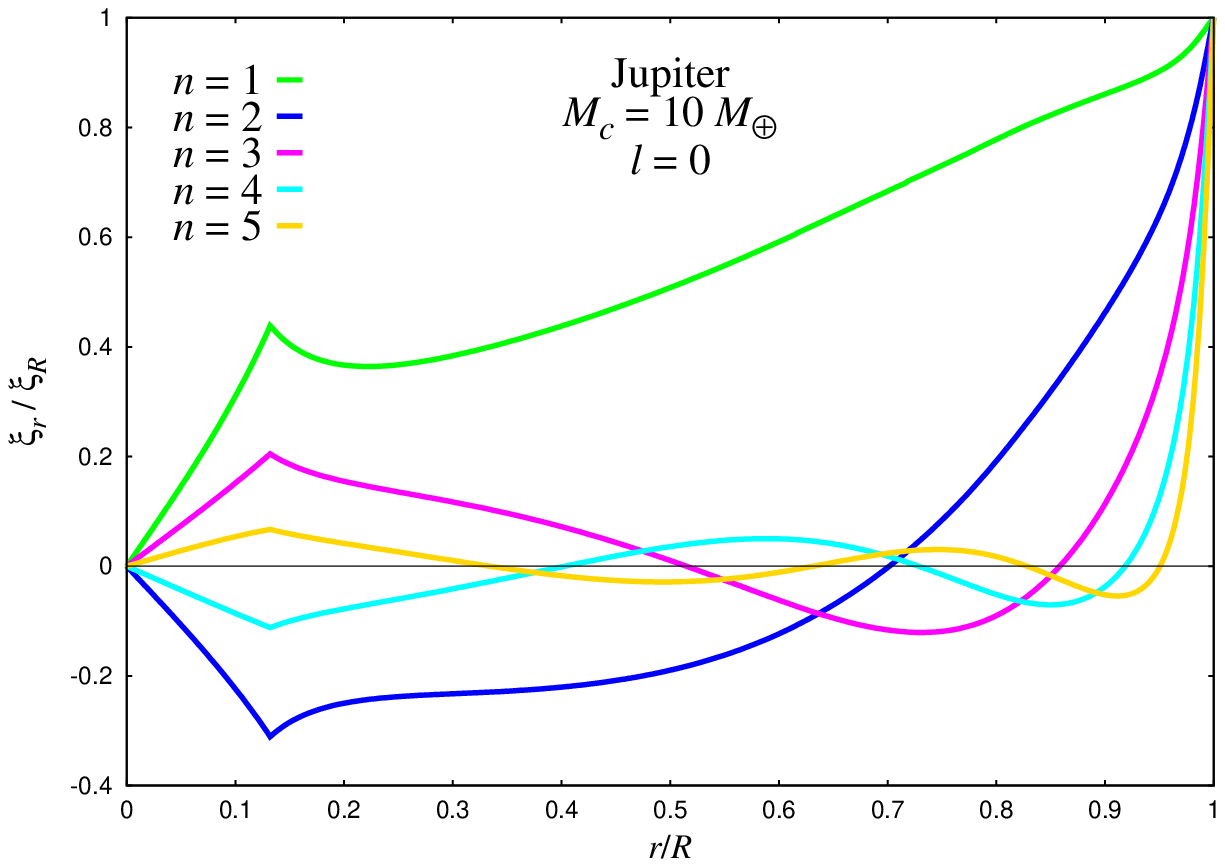}
\includegraphics[width = 0.5\textwidth]{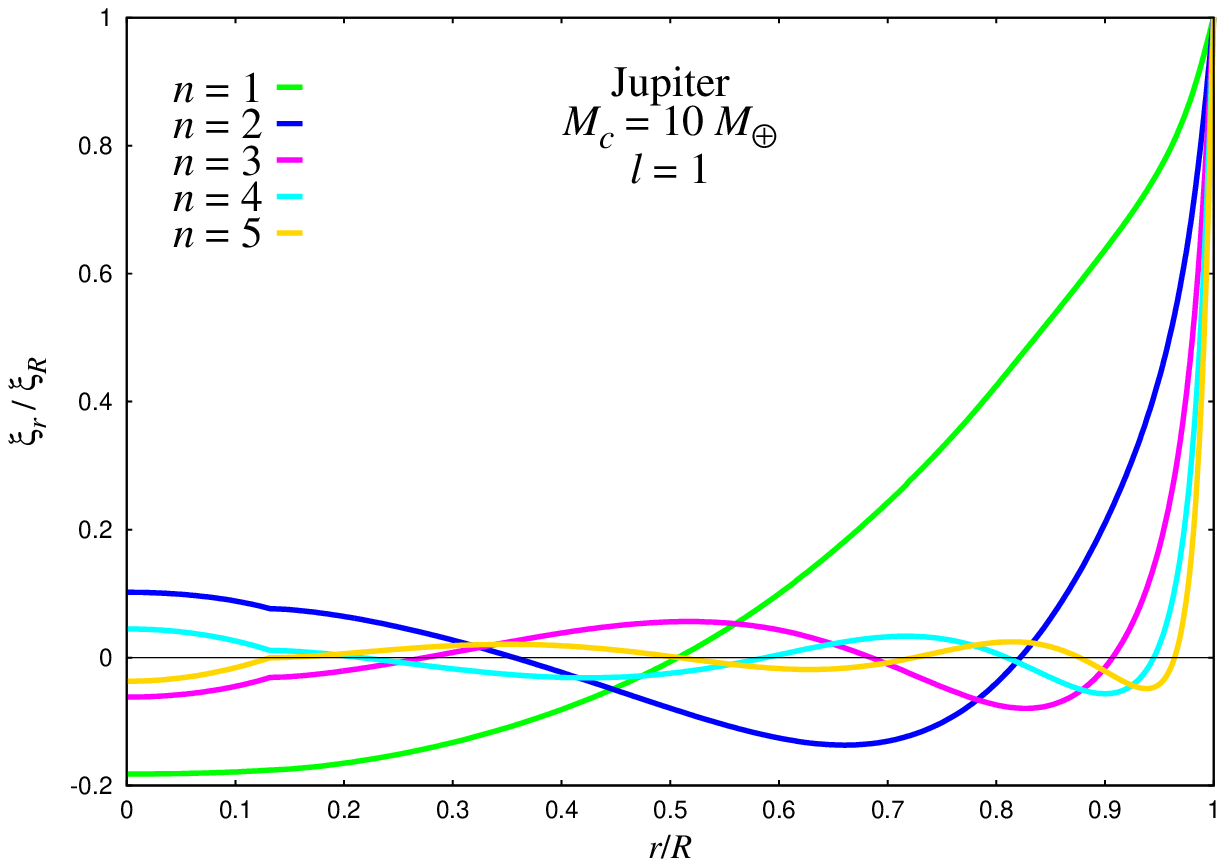}
\includegraphics[width = 0.5\textwidth]{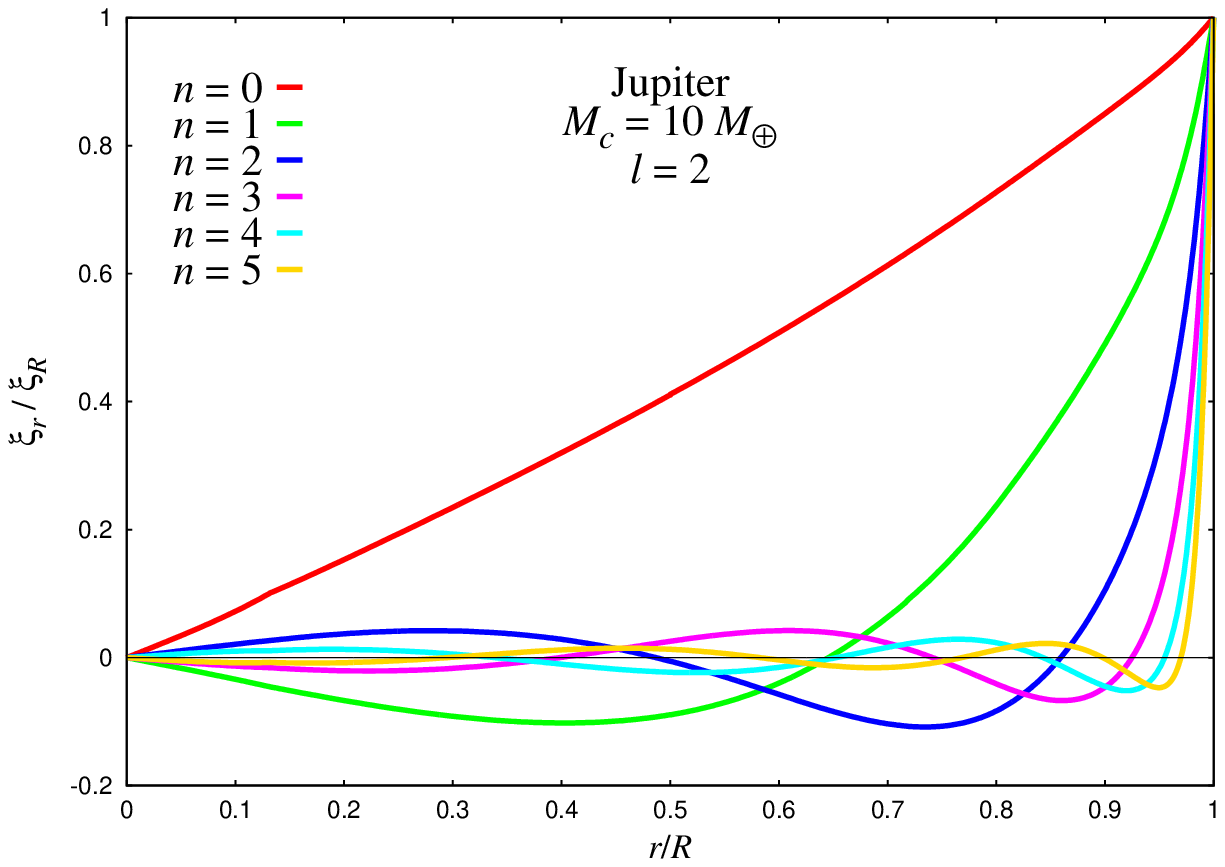}
\includegraphics[width = 0.5\textwidth]{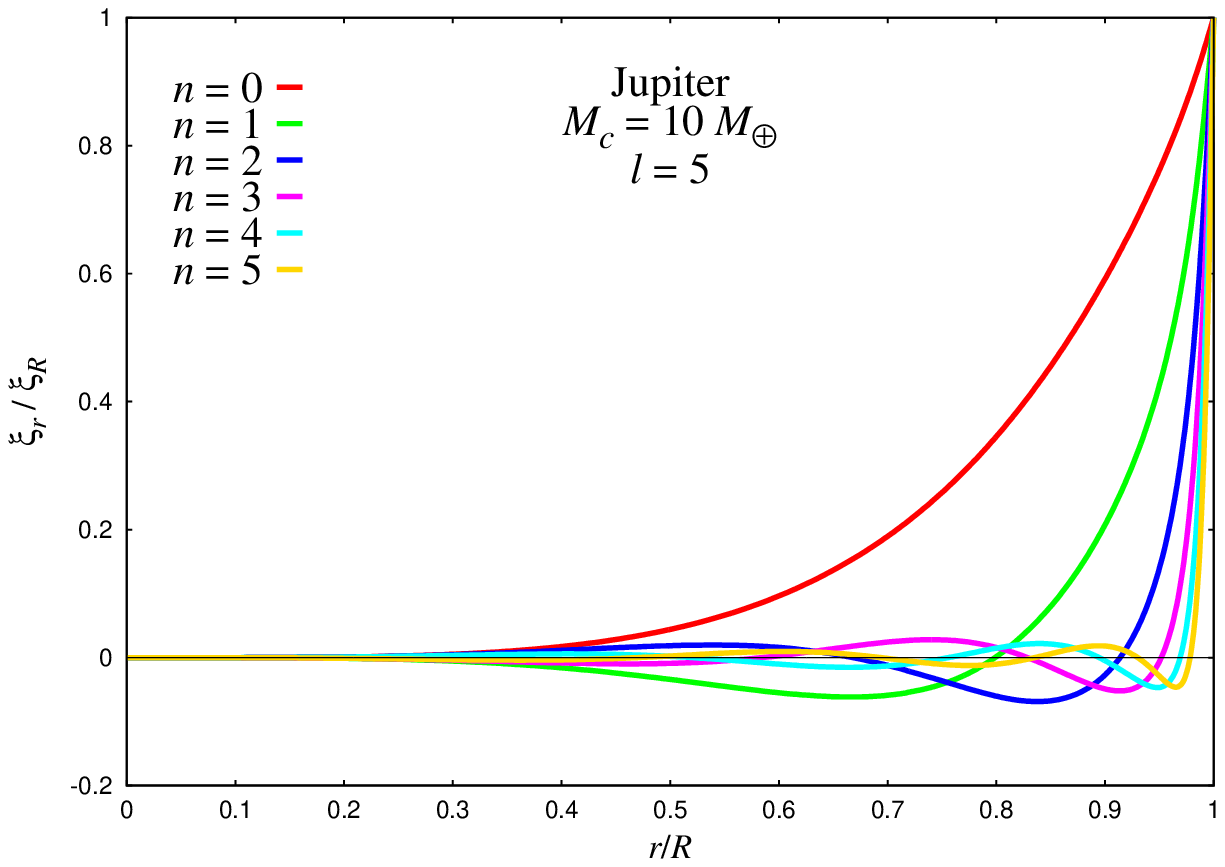}
\caption{Radial component of the eigendisplacement $\xi_r$ for low-degree, lowest-order modal oscillations of the model J4 of Jupiter, which has a core mass equal to $M_c = 10 M_{\oplus}$. The radial displacement is taken equal to 1 m at the surface; in other words, the radial displacement $\xi_r$ is normalized by its value $\xi_R$ at the surface. Top left: $l=0$; top right: $l=1$; bottom left: $l=2$; bottom right: $l=5$.}
\label{f3}
\end{figure}

Figure \ref{f3} portrays the radial component of the eigendisplacement $\xi_r$ for low-degree, lowest-order modal oscillations of the model J4 of Jupiter, as a function of the relative radius $x=r/R$. The radial displacement is shown for $l=0$, 1, 2, and 5. It is taken equal to 1 m at the surface, for every mode. The behavior of the modes near the center is determined by the boundary conditions of the specific numerical problem considered here \citep{unno89}, which lead to the following relations, for $ r \sim 0$:
\begin{eqnarray}
\xi_r & \sim & 0  \quad \quad \quad \text{ \ for  }  l = 0 \ , \nonumber \\
\xi_r & \propto & \left (\frac{r}{R} \right )^{l-1}  \text{  for  }  l \geq 1 \ .
\end{eqnarray}
Thus, for $l=1$, the radial displacement does not necessary vanish near the center. In Figure \ref{f3},  for the lowest degrees $l$, the presence of the dense core is directly visible at the boundary with the envelope ($x = 0.13$, for this model). For higher degree (here, $l=5$), the influence of the core is less obvious in the radial eigenfunctions, because the amplitude of the radial displacement vanishes near the center, for every radial order $n$.

\begin{figure}
\plotone{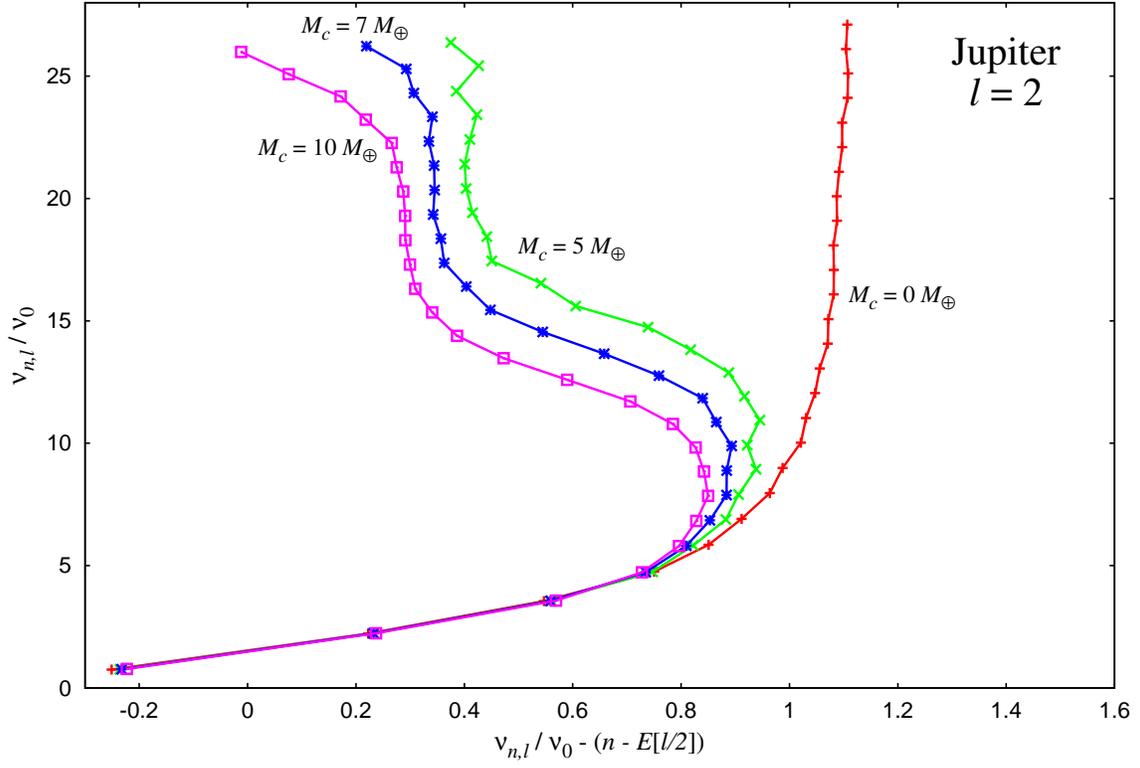}
\caption{Echelle diagrams of the Jovian eigenfrequencies calculated for different core masses with $\nu_0 = 155$ $\mu Hz$ and $l = 2$ (In Earth units, the corresponding core masses  are
$M_{core} = 0$: $+$,  $M_{core} = 5$: $\times$, $M_{core} = 7$: $\ast$, $M_{core} = 10$: $\boxdot$). \label{f4}}
\end{figure}

The influence of the size of the core on the frequency spectrum of Jupiter has already been studied \citep{provost93, gudkova99}. However, no determination of the derivative of the eigenfrequencies with respect to the core mass has yet been provided. Focusing on the f-modes and p-modes of Jupiter, we calculate their eigenfrequencies for various core masses, with the helium mass fraction fixed at $0.25$ in the envelope. An \'echelle diagram of the eigenfrequencies of Jupiter for $l=2$ and for a few core masses is given in Figure \ref{f4}. The spectra are very well separated for radial order $n \geq 4$,  which indicates, as mentionned by \citet{vorontsov89} and \citet{gudkova99}, that the high-frequency acoustic oscillations of low degree $l$ can be very useful in determining the structure and size of the core.

\begin{figure}
\includegraphics[width = 0.5\textwidth]{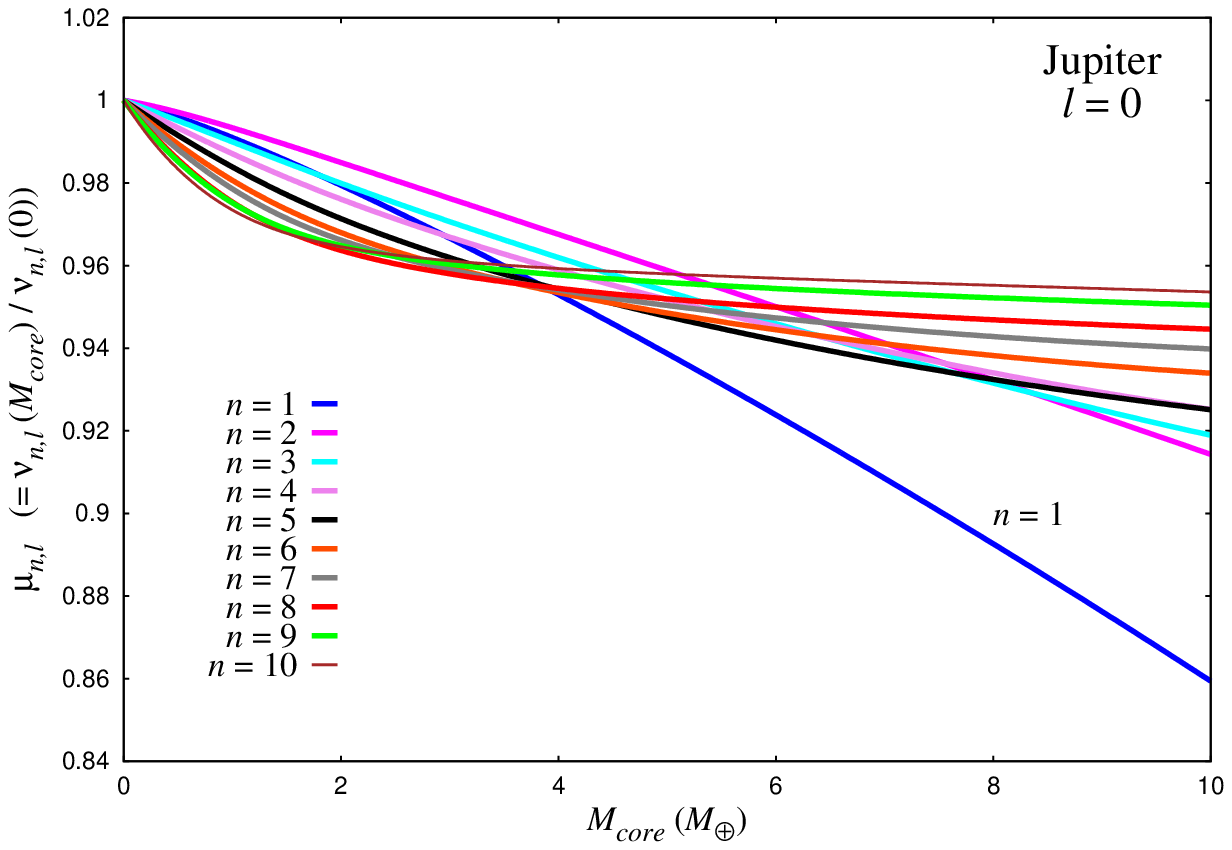}
\includegraphics[width = 0.5\textwidth]{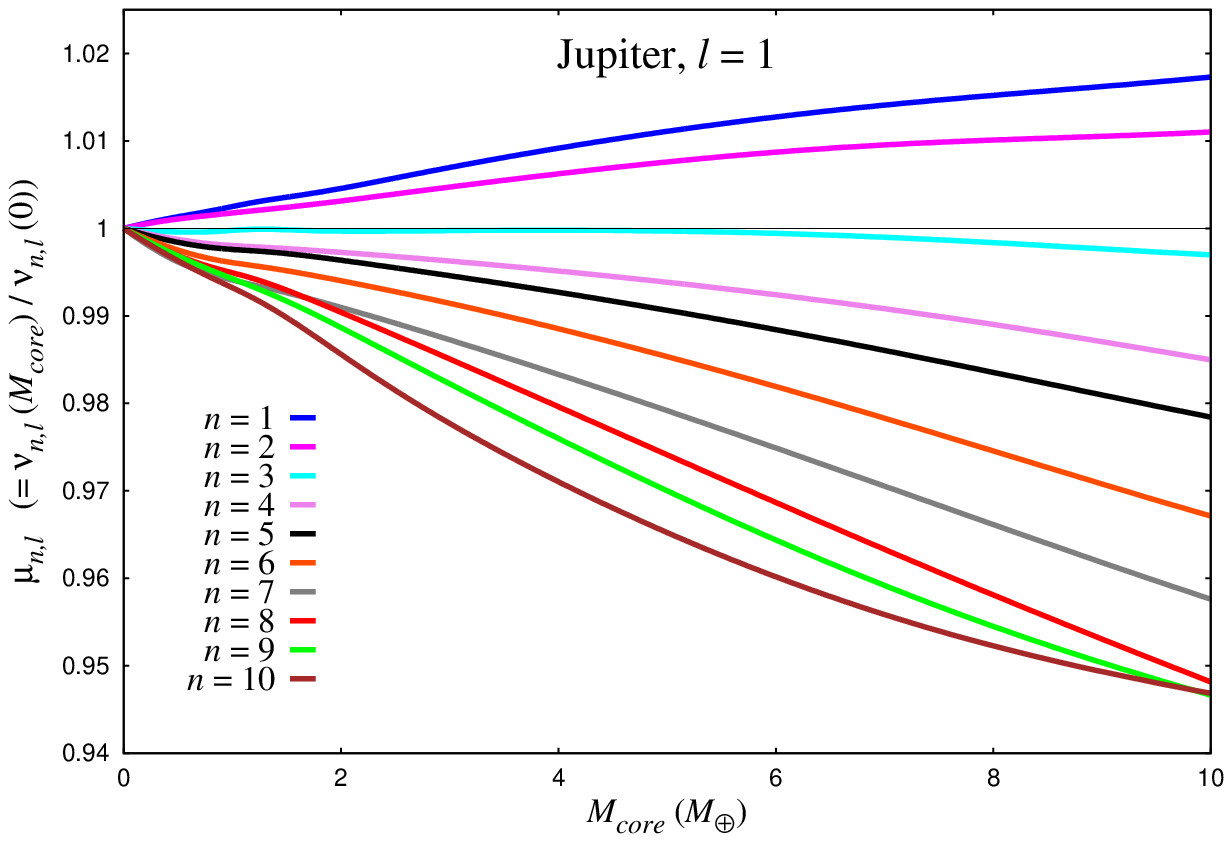}
\includegraphics[width = 0.5\textwidth]{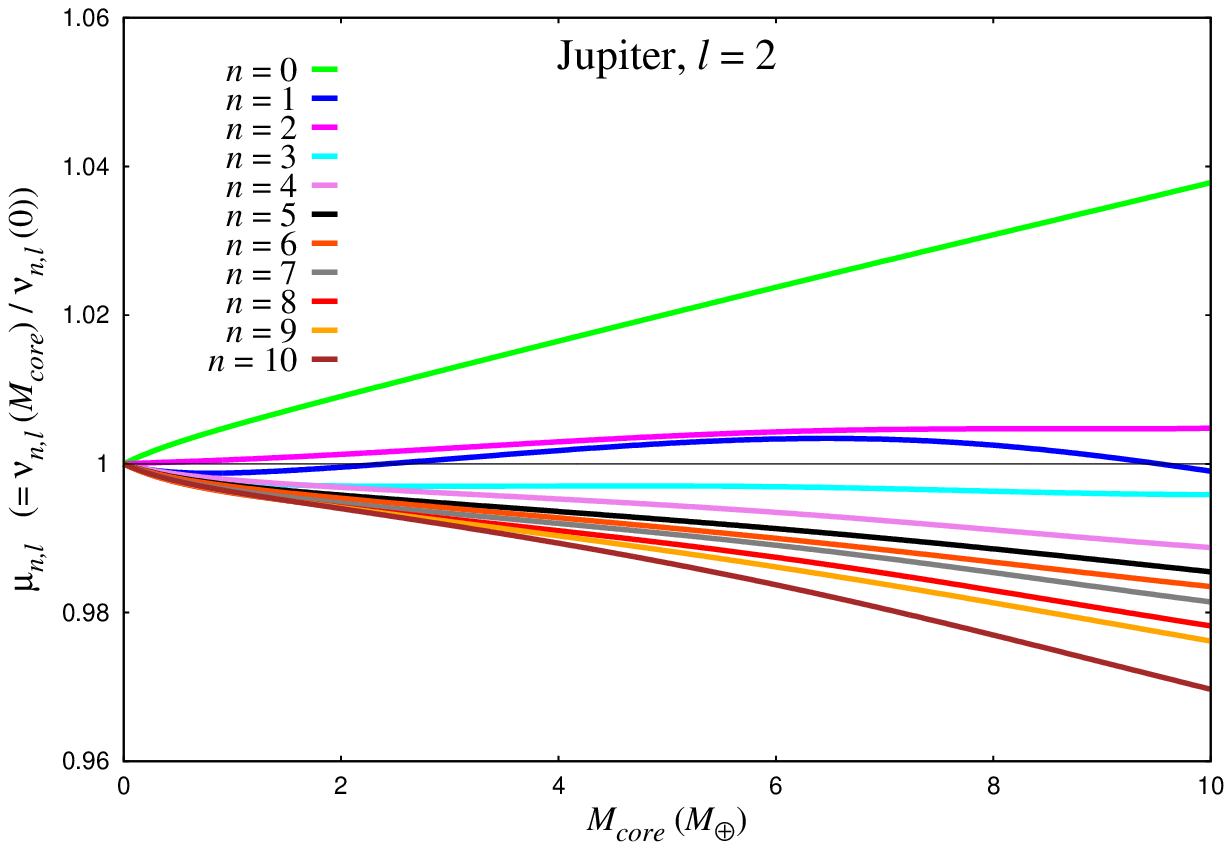}
\includegraphics[width = 0.5\textwidth]{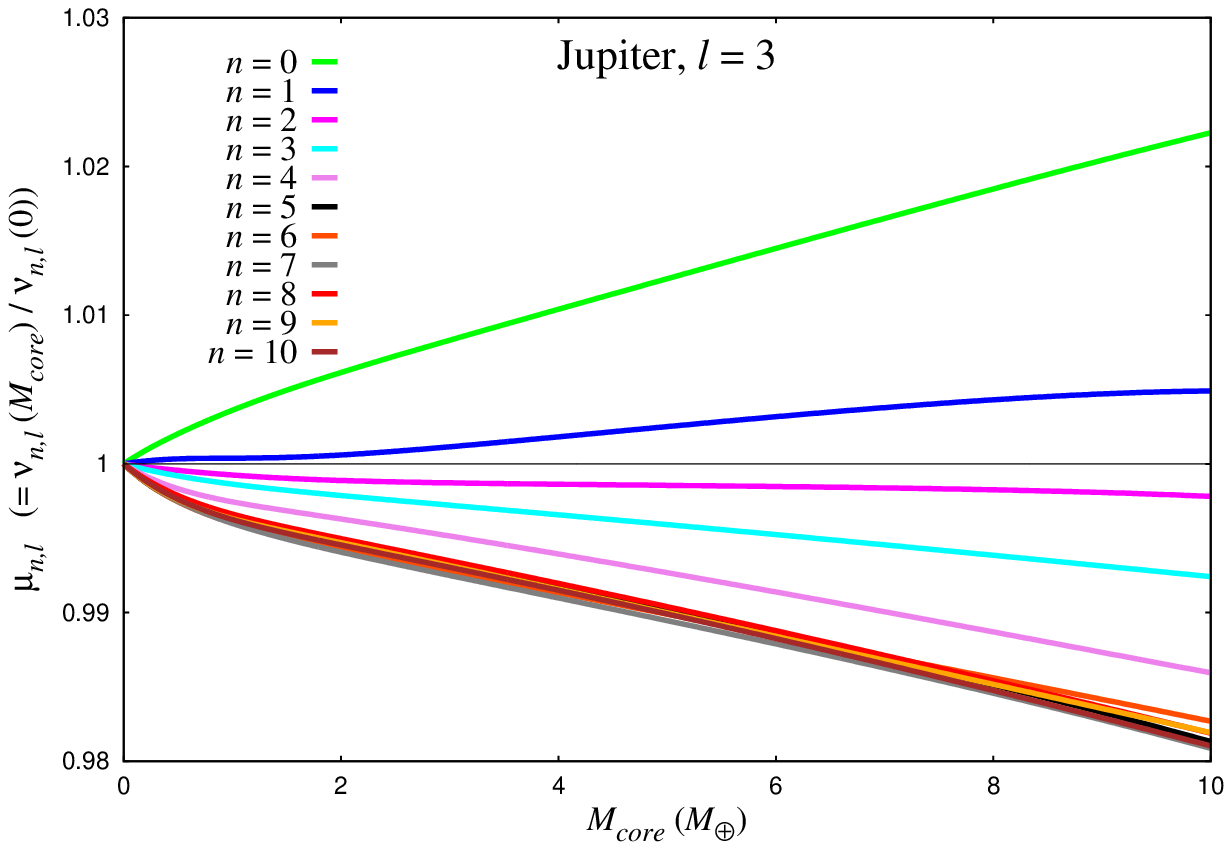}
\caption{Top left panel: Eigenfrequencies of p-modes for Jupiter as a function of the core mass for $l=0$. For every radial order $n$, the eigenfrequencies have been normalized by their coreless value: $\mu_{n,l}(M_{core}) = \nu_{n,l}(M_{core})/\nu_{n,l}(0)$. We assume that $0 \leq M_{core} \leq 10 M_{\oplus}$. Top right panel: same as for the top left panel, but for $l=1$. Bottom left panel: same as for the top left panel, but for $l=2$. Bottom right panel: same as for the top left panel, but for $l=3$.  \label{f5}}
\end{figure}

We highlight the sensitivity to the core mass of the eigenfrequencies of low-degree acoustic oscillations. Figure  \ref{f5} shows the eigenfrequencies of such modes for Jupiter as a function of the mass of the core, $M_c$, for $n \in [0,10]$ and for $l = 0$, $1$, $2$ and $3$. For every radial order $n$, the eigenfrequencies have been normalized by their coreless value:
\begin{equation}
\mu_{n,l}(M_{core}) = \frac{\nu_{n,l}(M_{core})}{\nu_{n,l}(0)} \ .
\end{equation}
This normalization allows us to compare the deviation of the frequencies from their coreless values as the core mass increases, regardless of their absolute value, which depends of the radial order $n$. 

For $l=0$, as $n$ increases the frequency becomes less sensitive to the size of the core. The normalized frequency $\mu_{1,0}$ is by far the most affected by the variation of the core mass; its value decreases by more than 14\% between the coreless version of Jupiter and the model with a $10$- Earth mass core. For $l \geq 1$, the trend is very different; the derivative of the normalized frequency with respect to the core mass decreases from positive to negative values as $n$ increases. As a result, this derivative approximately vanishes for specific frequencies, for example $\mu_{3,1}$ (Fig.\ref{f5}, top right), $\mu_{2,2}$ (Fig.\ref{f5}, bottom left), and $\mu_{2,3}$ (Fig.\ref{f5}, bottom right). All these frequencies vary by less than 0.5\% over a core mass range from 0 to 10 $M_{\oplus}$. For higher radial order, the influence of the core mass is more important; for example, $\mu_{10,2}$ varies by more than 3.0\% over the core mass range from 0 to 10 $M_{\oplus}$. This increased sensitivity is consistent with the previous discussion concerning the \'echelle diagram. The f-mode is also sensitive to the core mass, with a variation up to 3.8\% with core mass from 0 to 10 $M_{\oplus}$, for $l = 2$ and 3. These variations may be surprising compared to the stellar case, where the f-modes are constrained to shallow depths, far away from the influence of the core. However, this phenomenon has already been pointed out by \citet{marley91} in the Saturnian case.  On the other hand, planetary f-modes are also likely to decay with depth and, as the degree $l$ increases, the oscillation moves to the surface of the planet and its frequency is determined by the outer layers \citep{vorontsov76, gudkova99}. Calculations for higher degrees $l$ may thus exhibit a reduction of the influence of the core mass on the frequencies of the f-modes.
 We conclude, however, that for the low-degree nonradial oscillations the sensitivity of the pulsation frequencies to the core mass is important both for the f-modes and for the high-radial order p-modes. However, for some specific intermediate values of radial order $n$, the sensitivity nearly vanishes over the whole core mass range. If detected, these particular modes would provide little evidence of the presence of a core.

%%%%%%%%%%%%%%%%%%%%%%%%%%%%%%%%%%%%%%%%%%%%%%%%%%%%%%%%%%%%%%%%%%%%%%%%%%%%%%%%%%%%%%%%%%%%%%%%%%%%%%%%%%%%%%%%%%%%%%%%
%%%%%%%%%%%%%%%%%%%%%%%%%%%%%%%%%%%%%%%%%%%%%%%%%%%%%%%%%%%%%%%%%%%%%%%%%%%%%%%%%%%%%%%%%%%%%%%%%%%%%%%%%%%%%%%%%%%%%%%%

\section{The giant exoplanets}

\subsection{General description}

For a systematic look at the exoplanets currently known, we use the catalog developed by \citet{schneider11} and available at the URL \emph{http://www.exoplanet.eu}. As of the $7^{th}$ of July, 2012, $777$ confirmed planets are listed in this catalog. We limit our study to the planets whose radius and mass have both been estimated. Furthermore, we focus on the giant exoplanet regime and, therefore, we select the planets whose radii are within the range $0.8 \ R_J \lesssim R_P \lesssim 2.1 \ R_J$. In terms of mass, most of the detected giant exoplanets have masses less than 5 $M_J$, but the distribution has a long tail towards masses larger than 10 $M_J$ \citep{udry07}.  Numerically, 88\% of the selected exoplanets have masses less than or equal to 5 $M_J$, and 94\% have masses less than or equal to 10 $M_J$. In the 10 to 20 $M_J$ interval, it is difficult to fix a clear upper limit for giant exoplanets masses, because the planet population and the brown dwarf population overlap \citep{udry07, leconte09}. Therefore, though we restrict the mass range studied, we are aware of the ambiguous status of the heaviest objects. This final set is composed of $174$ exoplanets. 

We calculate the characteristic frequency $\nu_0$ for each object of this group, using the techniques and models developed and described in sections 2 and 3. The helium mass fraction is fixed at 0.25 in the entire envelope and no core is added. The results are shown in Table \ref{nu0exoplanets1}, with the planets sorted from low to high mass. For the selected objects, the characteristic frequency range is $33 \ \mu Hz \leq \nu_0 \leq 815 \ \mu Hz$. Around 1 $M_J$, $\nu_0$ is smaller than Jupiter's value for almost every object, since their radii are larger than 1 $R_J$. The spread of values is dramatic around every mass within the range [0.17 $M_J$, 30 $M_J$], which emphasizes the strong sensitivity of $\nu_0$ to planet parameters.

\begin{figure}
\plotone{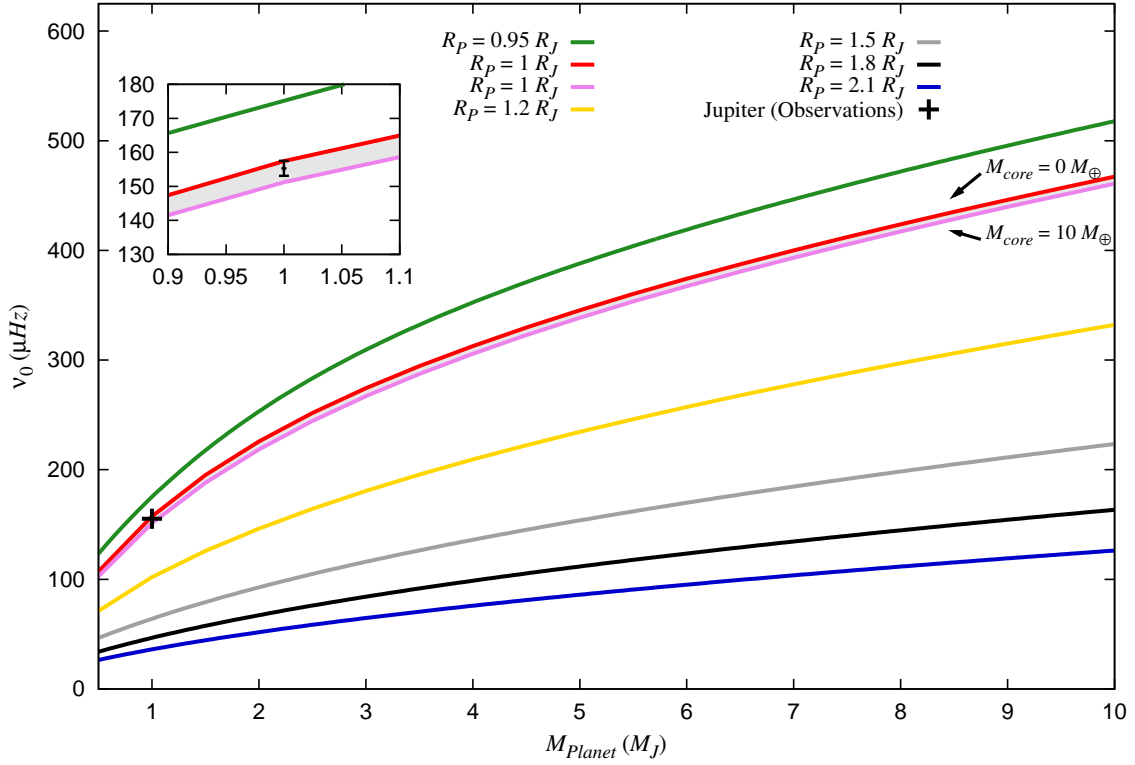}
\caption{The charateristic frequency, $\nu_0$, as a function of the planet mass, for various planet radii. The helium mass fraction  in the envelope has been set to 0.25. The models are all coreless except for $R_P$ = 1.0 $R_J$. For this radius value, we calculate the function $\nu_0(M_P)$ for various core masses between 0 $M_{\oplus}$ (solid red line) and 10 $M_{\oplus}$ (solid purple line). The figure includes an insert which zooms into the range 0.9 $\leq$ $M_P$ $\leq$  1.1 $M_J$. The gray area in the insert depicts the frequency range reached by $\nu_0$ for the models with $R_P$ = 1.0 $R_J$ and 0 $\leq$ $M_{core}$ $\leq$  10 $M_{\oplus}$. The observed point for Jupiter has been added, both on the figure (black cross) and on the insert (with errorbars). The value of its characteristic frequency is taken from the measurements of \citet{gaulme11}. \label{f6} }
\end{figure}

To investigate the crossed dependence of $\nu_0$ on the radius and the planet mass, we calculate it for a wide range of radii and masses in the observed giant planet regime. Figure \ref{f6} portrays the corresponding results for 0.5 $\leq$ $M_P$ $\leq$  10 $M_J$, and 0.95 $\leq$ $R_P$ $\leq$  2.1 $R_J$. All the planet models are coreless, except for $R_P= 1.0$ $R_J$. For this radius value, we calculate the function $\nu_0(M_P)$ for several core masses within the Jovian range 0 $\leq$ $M_{core}$ $\leq$  10 $M_{\oplus}$. We place the observed point for Jupiter, taken from \citet{gaulme11},  at $M_P=1.0$ $M_J$. As can be seen, our model is consistent with the observational data. For any fixed value of the planet radius, it is clear that $\nu_0$ is an increasing function of the planet mass $M_P$. However, the sensitivity of $\nu_0$ to the planet mass decreases as the planet radius increases. In order to quantify this, we fit the curves of  Figure \ref{f6} with straight lines, in the high-mass regime (5 $\leq$ $M_P$ $\leq$  10 $M_J$), and calculate their derivatives. We find that, for a radius equal to 0.95 $R_J$, the corresponding derivative is 25 $\mu Hz$ $M_J^{-1}$. At the extreme opposite end of the giant planet radius spectrum, for a radius equal to 2.1 $R_J$, the corresponding derivative is 8.0 $\mu Hz$ $M_J^{-1}$. These conclusions are qualitatively consistent with the presumption that $\nu_0$ would scale approximately with the square root of the mean density of the planet \citep{jackiewicz12}. Thus, the asymptotic scaling relations between Jupiter and giant exoplanets would be roughly:
\begin{equation}
\nu_0 \sim \nu_{0,J}\left(\frac{M}{M_J}\right)^{0.5}\left(\frac{R}{R_J}\right)^{-1.5} \ , \label{nu0J}
\end{equation}
where $\nu_{0,J}$ refers to the mean frequency spacing of Jupiter.  As we show in \S\ref{mass} and \S\ref{42},
this expression only approximately holds.

Below, we investigate the separate influence of the radius, the mass, and the core mass on giant exoplanet pulsation modes. We focus on one parameter at a time. We select three specific quantities to discuss this influence: the characteristic frequency $\nu_0$, the eigenfrequencies of low-degree f-modes, and the eigenfrequencies of low-degree p-modes.
To determine the influence of planetary parameters on the characteristic frequency and on the frequency spectrum of exoplanets, we calculate these for a wide range of each parameter (radius, mass, entropy, and core mass), all other things being equal.

%%%%%%%%%%%%%%%%%%%%%%%%%%%%%%%%%%%%%%%%%%%%%%%%%%%%%%%%%%%%%%%%%%%%%%%%%%%%%%%%%%%%%%%%%%%%%%%%%%%%%%%%%%%%%%%%%%%%%%%%

\subsection{Eigenfrequencies and characteristic frequencies of giant planets}

\subsubsection{Dependence on the planet mass}
\label{mass}

\begin{figure}
\includegraphics[width = 0.5\textwidth]{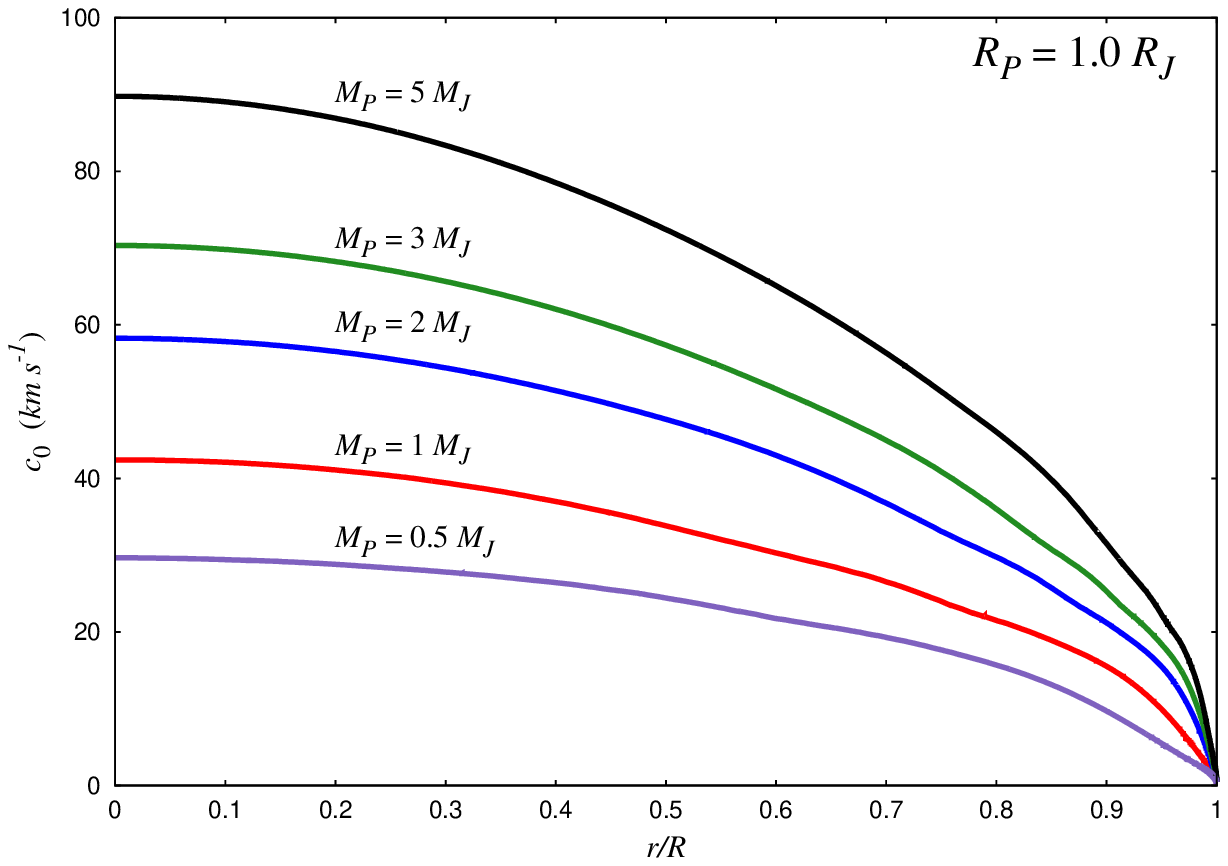}
\includegraphics[width = 0.5\textwidth]{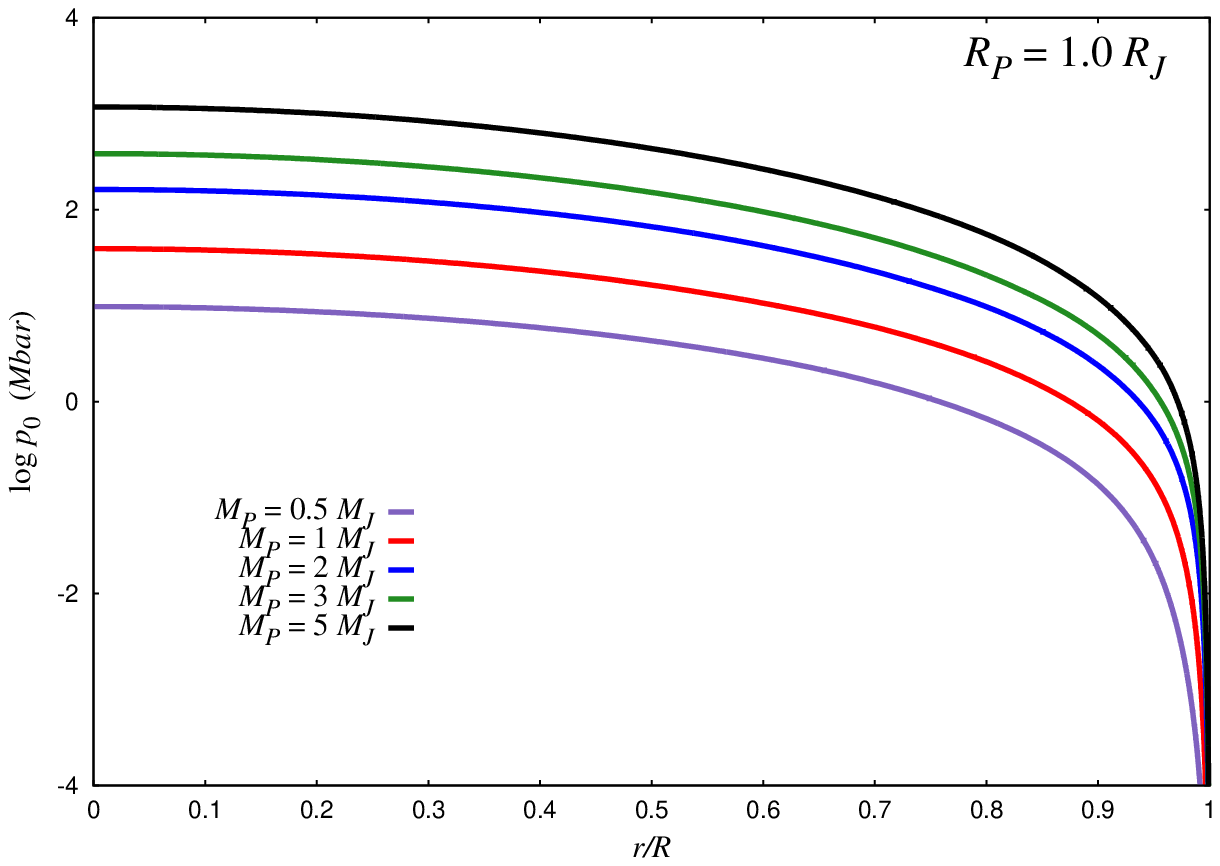}
\caption{Left panel: distribution of sound speed $c_0$ $(km \ s^{-1})$ as a function of the
relative radius $r/R$ for coreless exoplanet models with various masses, with the radius fixed at $R_P = 1.0$ $R_J$. 
Right panel: distribution of pressure $p_0$ $(Mbar)$ for the same models, with a logarithmic scale for the Y-axis. 
At every radius $r$, both the sound speed and the pressure are increasing functions of the planet mass, when the planet radius is fixed at $R_P = 1.0$ $R_J$.
\label{f7}
}
\end{figure}

We build several planetary models with the radius fixed at $R_P = 1.0$ $R_J$ and with various masses. We use coreless models, since we are here exploring the dependence on mass. Table \ref{parametersMass} presents various parameters of these models: the planet mass, $M_P$ (in Jupiter units),  the central pressure, $p_c$ (in Mbars), the central density, $\rho_c$ (in cgs units), the specific entropy, $S$  (in $k_B  / baryon$), and the characteristic frequency, $\nu_0$ (in $\mu Hz$). When the radius is fixed at 1.0 $R_J$, $\nu_0$ is an increasing function of planet mass. We fit this function with a power law and obtain:
\begin{equation}
\nu_0(M_P) \sim 164.0 \times \left(\frac{M_P}{M_J}\right)^{0.48} \mu Hz \ . \label{nu0MP}
\end{equation}
Thus, we derive a power law consistent with the asymptotic scaling relation suggested by eq. \ref{nu0J}.
Figure \ref{f7} depicts the profiles of the pressure $p_0$ and the sound speed $c_0$ along the relative radius $x=r/R$, in hydrostatic equilibrium, for the first models of Table \ref{parametersMass}. As can be seen, at every level of the relative radius $x$, when the radius is fixed, both the pressure and the sound speed are increasing functions of the planet mass. Thus, given its definition (Eq. \ref{nu0}), $\nu_0$ increases as the planet mass increases, all other things being equal. 

According to the asymptotic theory, we know that, for a given low value of the degree $l$, and for large radial orders $n$, $\nu_0$ is approximately the frequency gap between two modes of consecutive radial order: $\nu_{n+1,l} - \nu_{n,l} \sim \nu_0$. Thus, at high frequency, the spacing between eigenfrequencies increases when the planet mass increases, for a given value of the planet radius. Numerically, when the radius is fixed at $R_P = 1.0$ $R_J$, for an object of 1.0 $M_J$, we know that the high-frequency modes are separated by $\sim$$155$ $\mu Hz$, since this corresponds to the Jupiter case. For a 5.0-$M_J$ planet, the frequency gap between high-frequency modes is $\sim$$350$ $\mu Hz$, and, at the end of the giant planet regime, for a planet mass of 15.0 $M_J$, the frequency gap exceeds 500 $\mu Hz$. 

\begin{figure}
\plotone{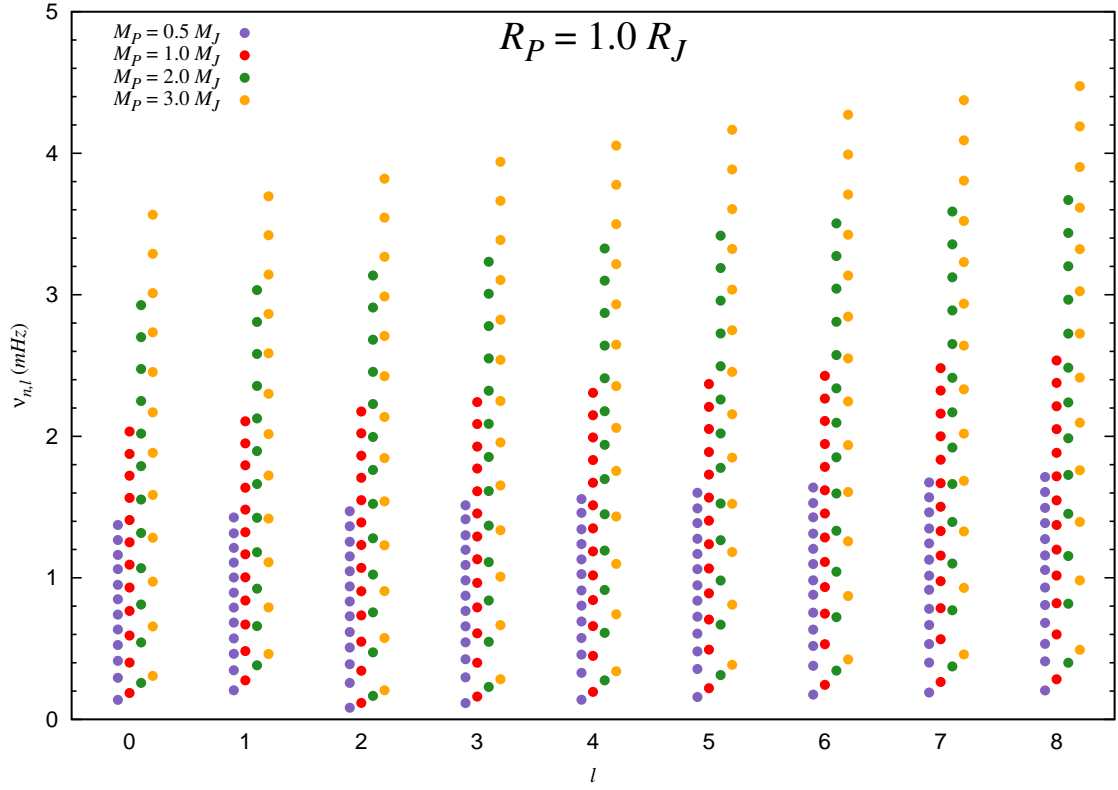}
\caption{Low-order eigenfrequencies of oscillation modes (in $mHz$) as a function of the degree $l$ for coreless exoplanet models with various masses, with the radius fixed at $R_P = 1.0$ $R_J$. The helium mass fraction in the envelope is 0.25. For the sake of clarity, the eigenvalue dots of the different models has been seperated and drawn on both sides of each integer degree $l$.  \label{f8} }
\end{figure}

We now calculate the eigenfrequencies of the lowest-order p-modes for the objects defined in Table \ref{parametersMass}, and for $l \in [0,8]$. Figure \ref{f8} presents the corresponding eigenvalues for $n$ up to 12, as a function of the degree $l$, for objects with mass equal to 0.5, 1.0, 2.0,  and 3.0 $M_J$, and a radius fixed at 1.0 $R_J$. The frequency spectra of the four planets appear more and more distinct from one another as we go up in frequency, and as we go up in degree $l$. Indeed, at low $l$, the f-modes of the four planets are close to one another, whereas the difference between the modes with the same $n$ and $l$ increases rapidly with frequency. Numerically, for $l=2$, the f-modes of the four planets are all within the range 0.08 $\leq$ $\nu_{0,2}$ $\leq$ 0.21 $mHz$. For $l=2$ and $n=12$, the difference between the eigenvalues for $M_P = 0.5$ $M_J$ and $M_P = 3.0$ $M_J$ is more than 2.3 $mHz$. This discrepancy at high radial order $n$ is, of course, due to the differences of the characteristic frequency $\nu_0$, which is a measure of the frequency scale at low-degree $l$ and high-order $n$. At high frequency, the value of $\nu_0$ for each planet is clearly visible on Figure \ref{f8}.

\begin{figure}
\includegraphics[width = 0.5\textwidth]{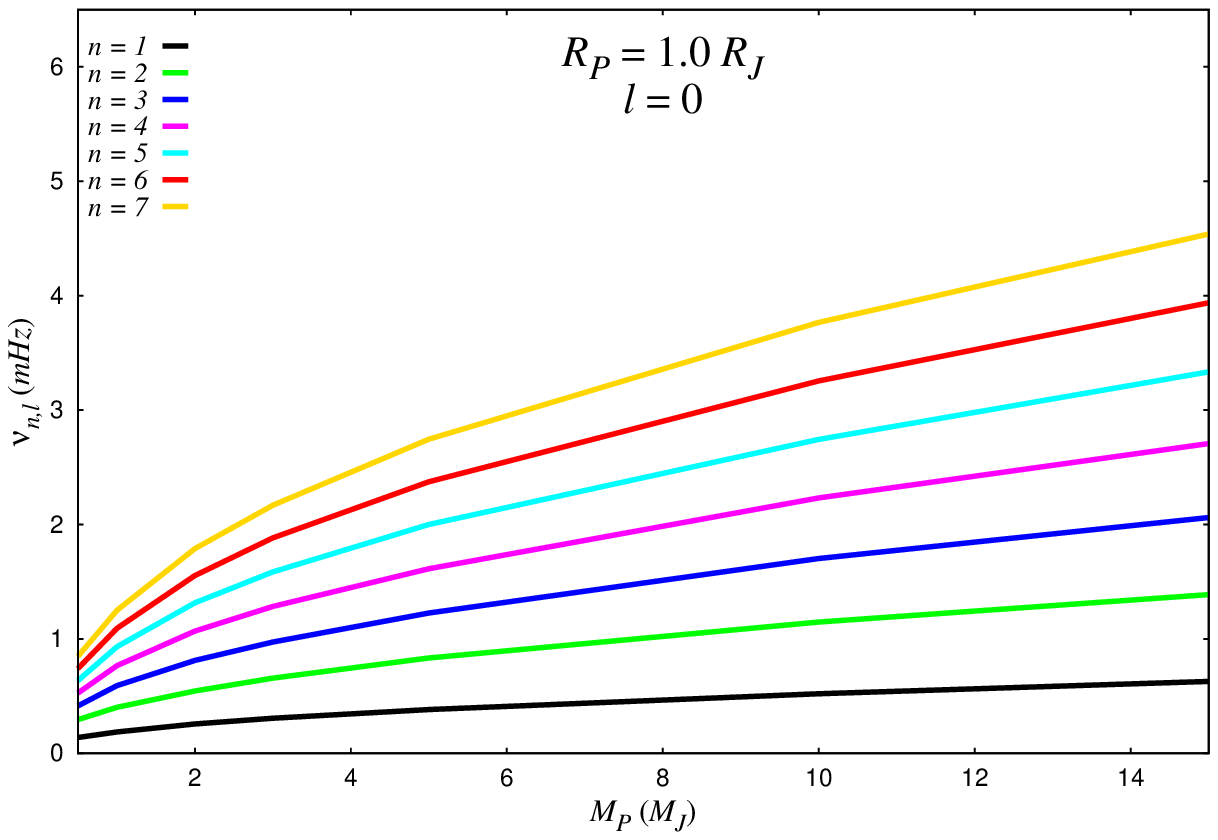}
\includegraphics[width = 0.5\textwidth]{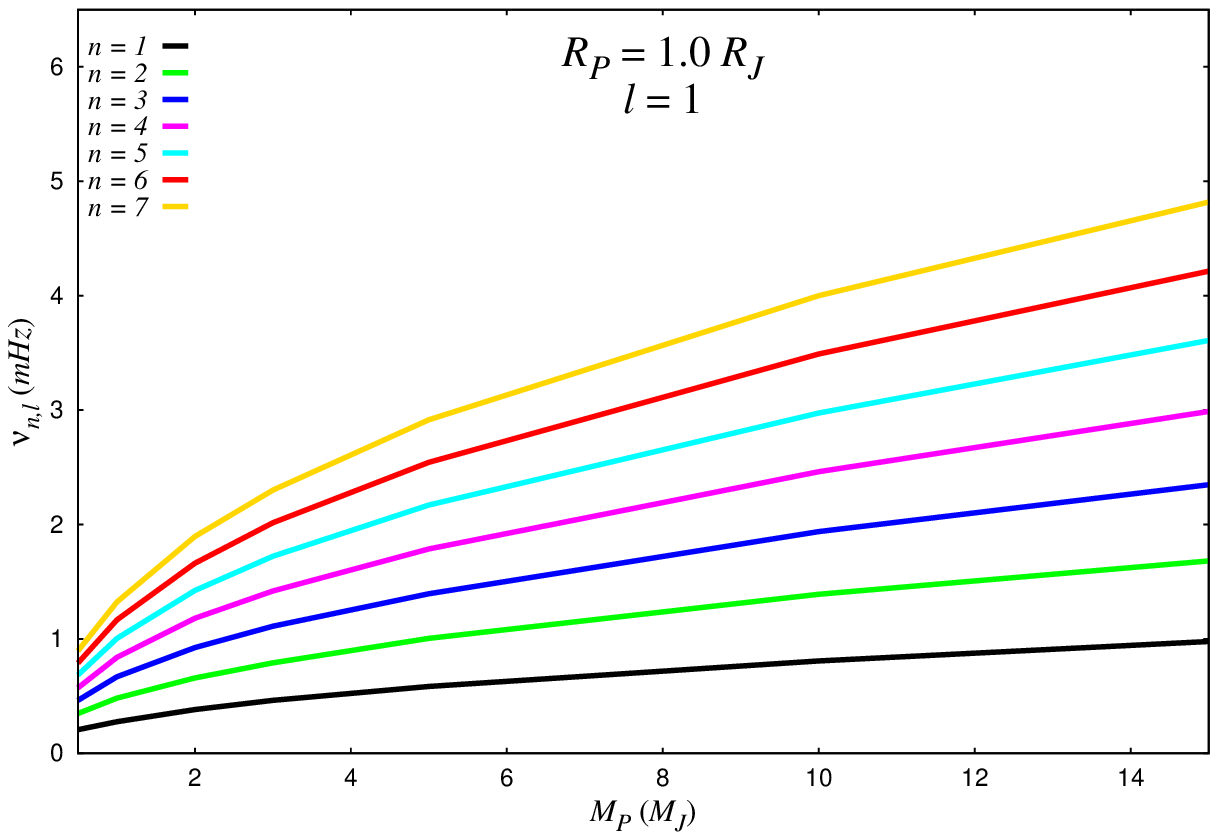}
\includegraphics[width = 0.5\textwidth]{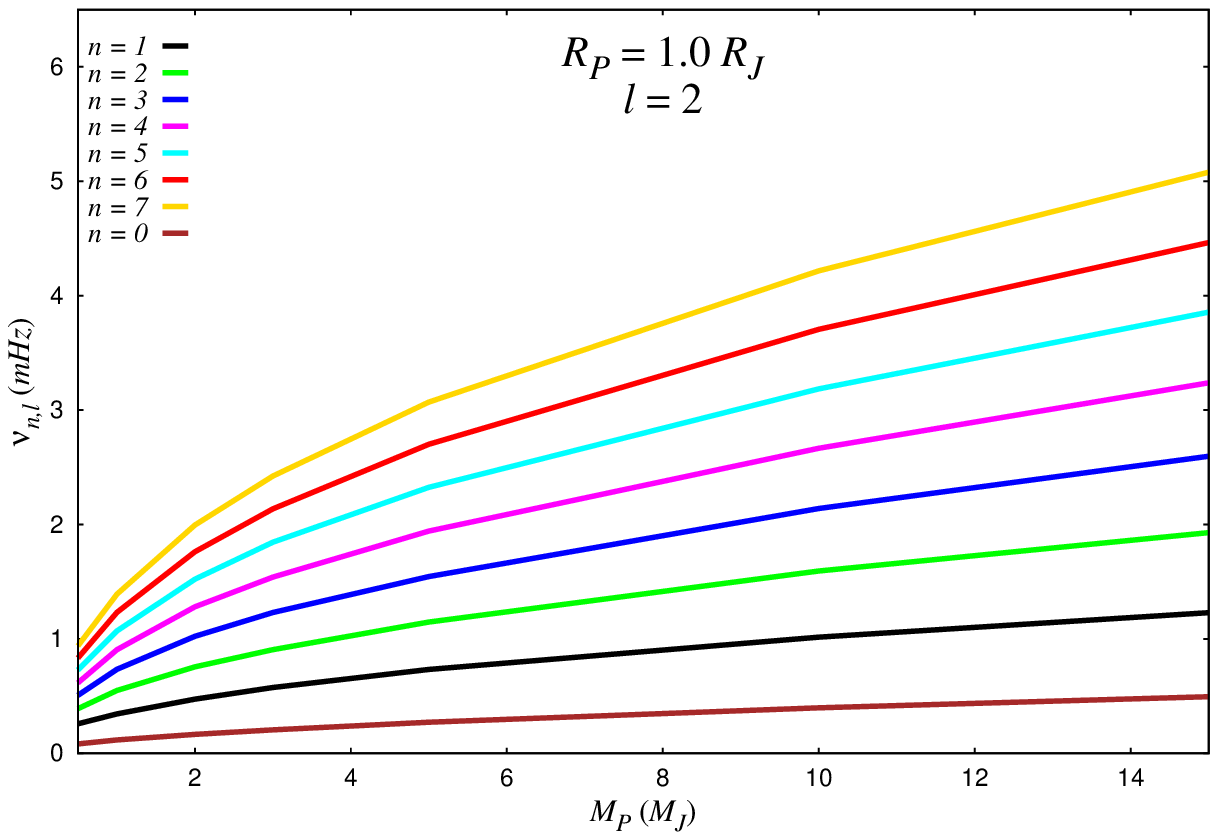}
\includegraphics[width = 0.5\textwidth]{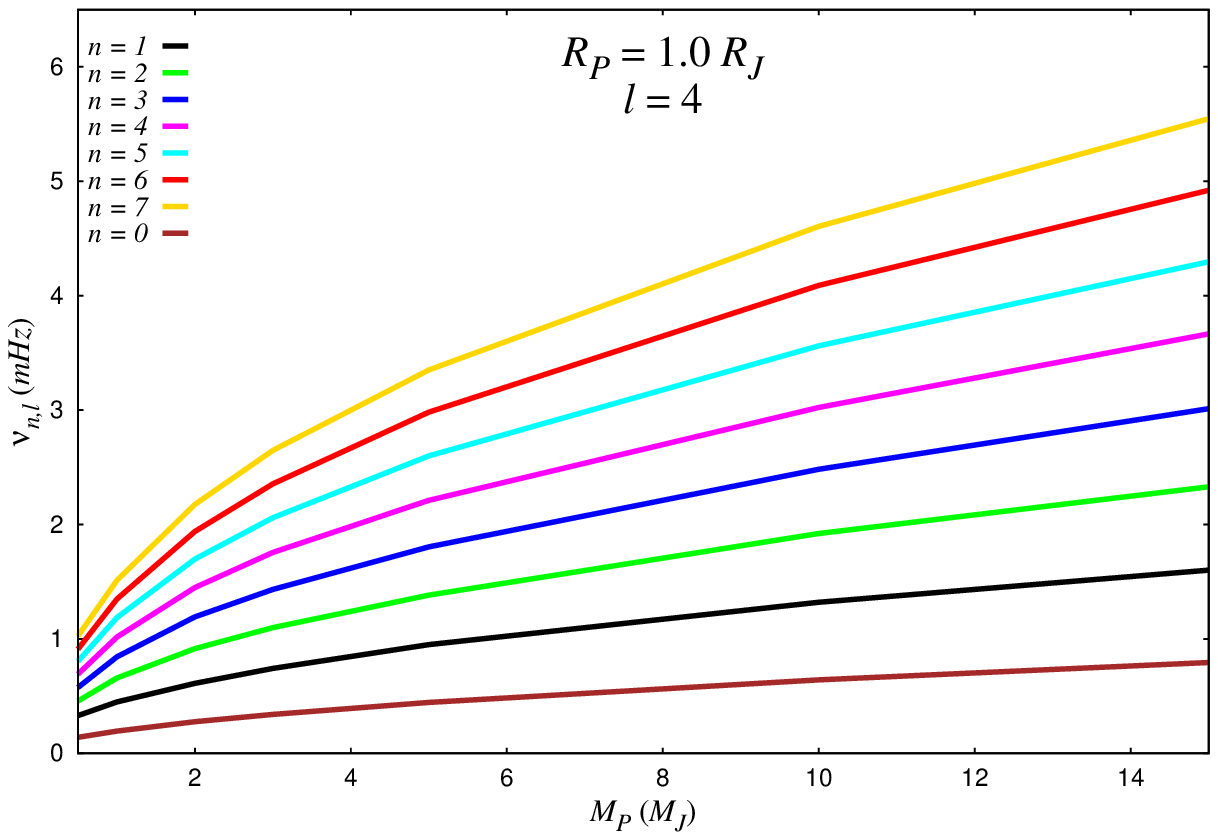}
\includegraphics[width = 0.5\textwidth]{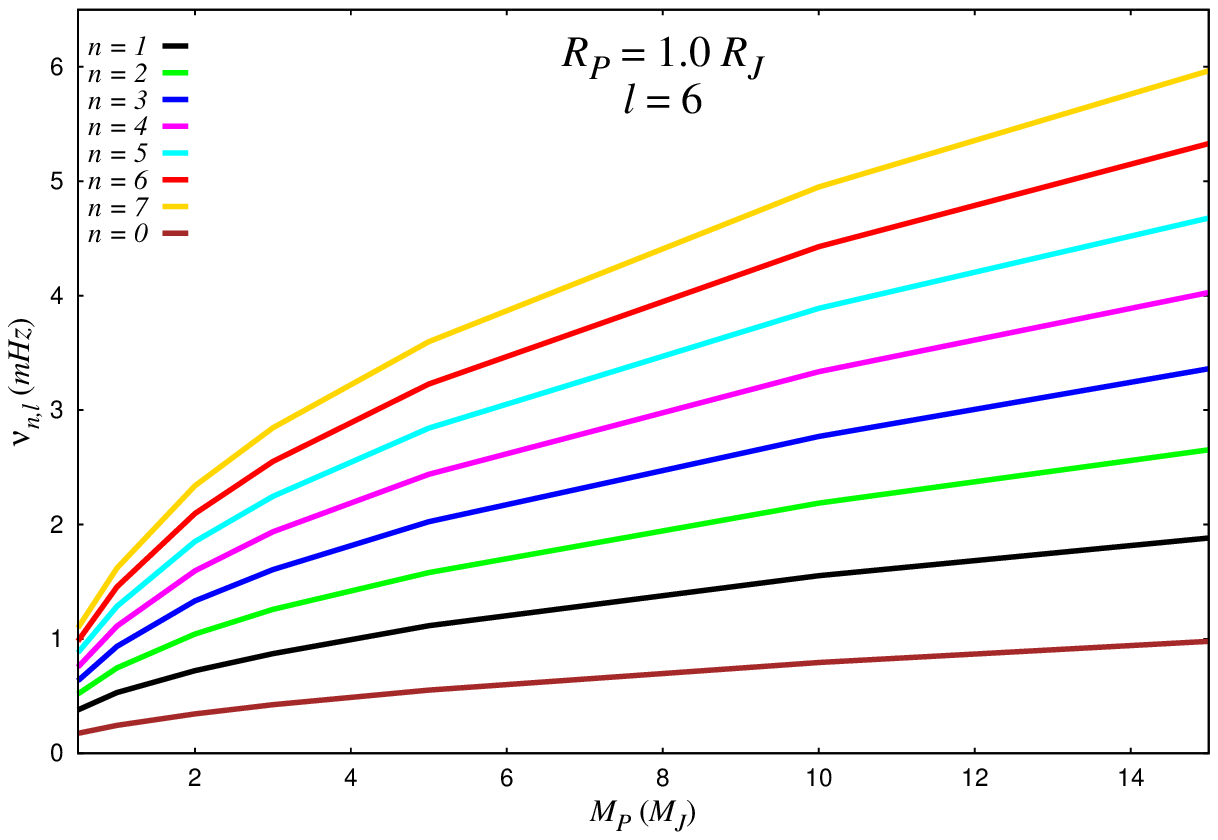}
\includegraphics[width = 0.5\textwidth]{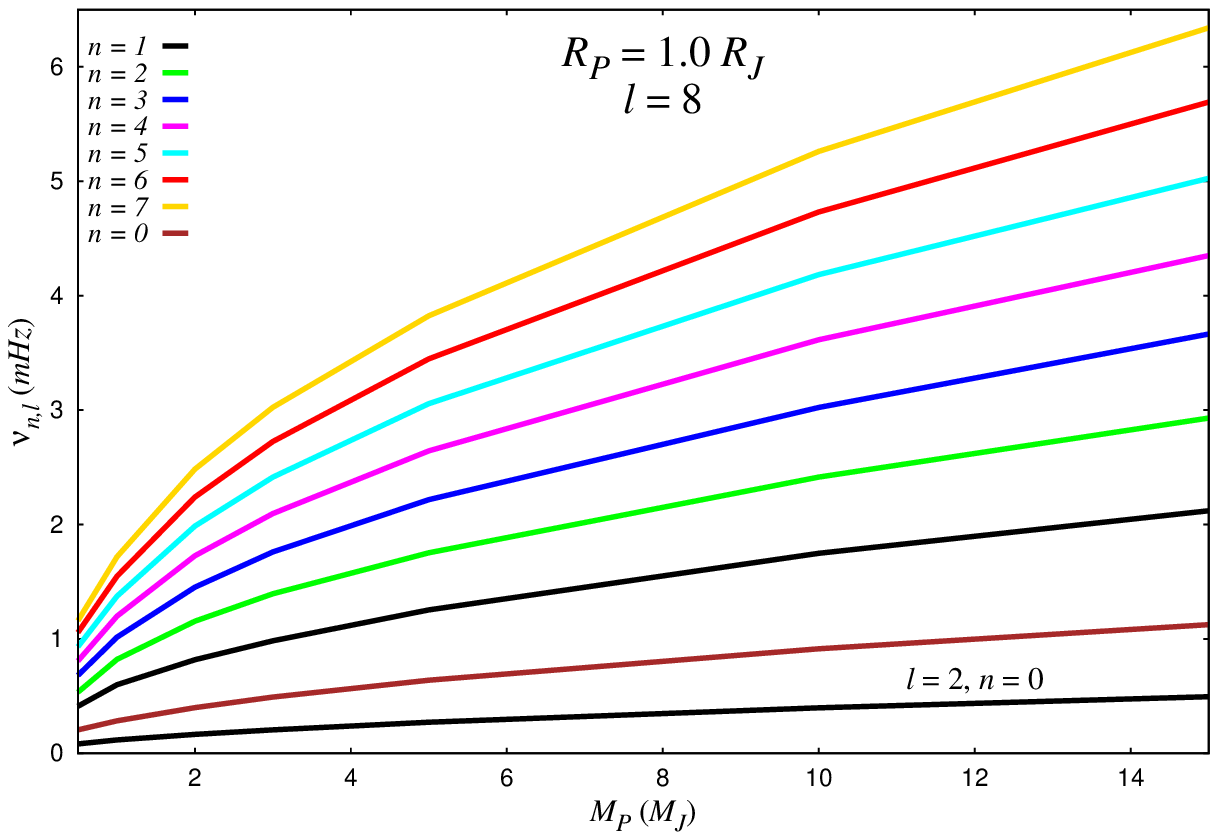}
\caption{Eigenfrequencies of oscillation modes (in $mHz$) as a function of the planet mass for coreless exoplanet models with the radius fixed at $R_P = 1.0$ $R_J$, and for various values of the degree $l$.
For the represented modes, for each value of the planet mass $M_P$, the frequency minimum and maximum are obtained for $(n,l) = (0,2)$ (middle left panel) and $(n,l) = (7,8)$ (bottom right panel), respectively.
On the bottom right panel ($l=8$), the functions $\nu_{0,2}(M_P)$ has been added (black dashed line). Thus, the frequency range of the calculated modes is contained within the solid gold line, defined by $(n,l) = (7,8)$,and the black dashed line, defined by $(n,l) = (0,2)$.
\label{f9}}
\end{figure}

This increase of the frequency range continues as the planet mass increases beyond 3.0 $M_J$. To appreciate the difference numerically, we focus on $l \in [0,8]$ and $n \in [0,7]$. Figure \ref{f9} presents several low-order eigenvalues, as a function of the planet mass, for various values of the degree $l \in [0,8]$. For the calculated modes, it appears from the calculations that the minimum in frequency is always obtained for $(n,l) = (0,2)$ (middle left panel), and the maximum is obtained for the highest $n$ and $l$ considered: $(n,l) = (7,8)$ (bottom right panel). This statement is true for every value of the planet mass $M_P$ in the selected range. On the bottom right panel ($l=8$), the functions $\nu_{0,2}(M_P)$ have been added (black dashed line). Thus, the frequency range of the calculated modes is contained between the black dashed line ($\nu_{0,2}$) and the solid gold line, defined by ($\nu_{7,8}$). Numerically, The low-degree, low-order eigenfrequencies of a 1.0-$M_J$ planet are in the range [0.11,1.8] $mHz$, whereas the same eigenfrequencies of a 15-$M_J$ object are in the range [0.50, 6.4] $mHz$.

\subsubsection{Dependence on the planet radius}
\label{42}

We build several planetary models with the mass fixed at $M_P = 1.0$ $M_J$ and with various radii. Table \ref{parametersRadius} presents various parameters of these models: the planet radius, $R_P$ (in Jupiter units),  the central pressure, $p_c$ (in Mbars), the central density, $\rho_c$ (in cgs units), the specific entropy, $S$ (in $k_B  / baryon$), and the characteristic frequency, $\nu_0$ (in $\mu Hz$). When the mass is fixed at 1.0 $M_J$, $\nu_0$ is a decreasing function of the planet radius. We again fit this function with a power law and obtain:
\begin{equation}
\nu_0(R_P) \sim 164.0 \times \left(\frac{R_P}{R_J}\right)^{-2.09} \mu Hz \  . \label{nu0RP}
\end{equation}
We find here a dependence on the radius slightly stronger than the one suggested by the asymptotic scaling relation of eq. \ref{nu0J}, though one has to keep in mind the discrepancy between the complexity of the interior models and the roughness of the relation exhibited in equation \ref{nu0J}.
If we compare Equations \ref{nu0MP} and \ref{nu0RP}, we see that, for the selected ranges of values, the dependence of $\nu_0$ on the radius is significantly more important than the dependence on the mass.
As explained in the previous subsection,  $\nu_0$ is approximately the frequency gap between two modes of consecutive radial order,  for a given low value of the degree $l$, and for large radial order $n$. Thus, this decreasing behavior results in a diminution of the frequency gap between high-frequency modes. Numerically, we can see that this gap is around $40$ $\mu Hz$ for a $2.0$-$R_J$ planet (again, the mass is equal to $1.0$ $M_J$), which is less than 26\% of the Jovian value.

\begin{figure}
\plotone{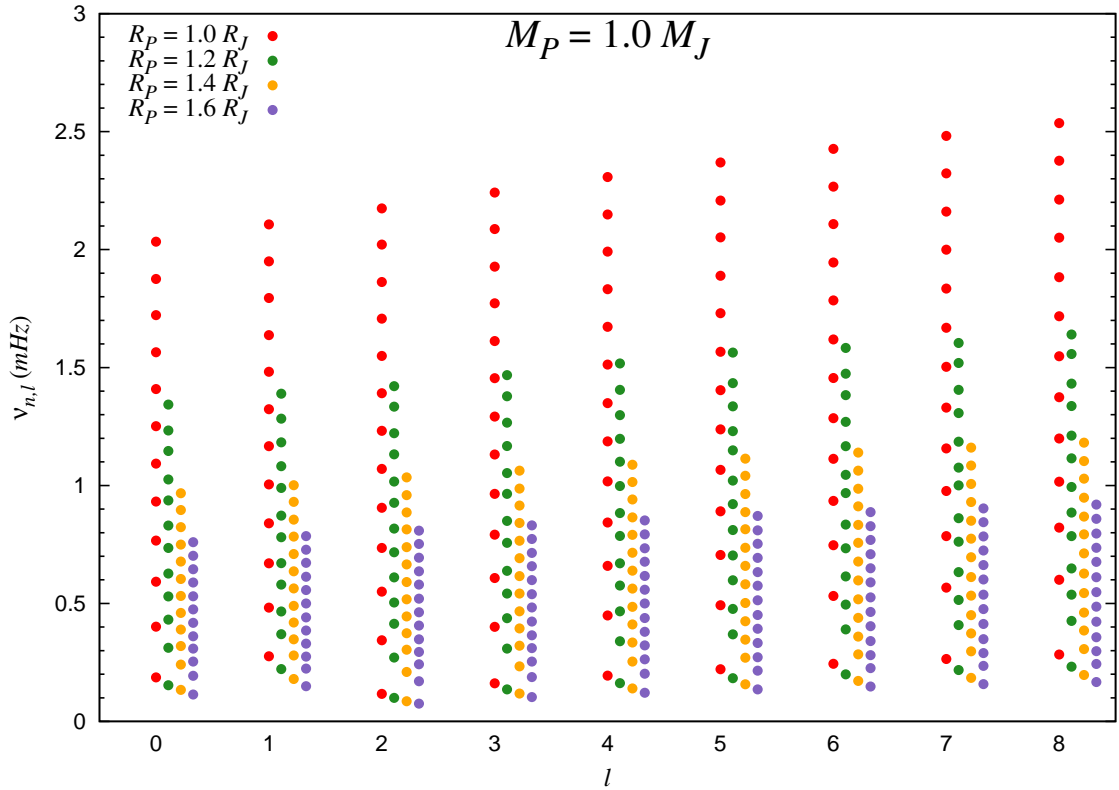}
\caption{Eigenfrequencies of oscillation modes (in $mHz$) as a function of the degree $l$ for exoplanet coreless models of various radii, with the mass fixed at $M_P = 1.0$ $M_J$. The helium mass fraction in the envelope is 0.25. For the sake of clarity, the eigenvalue dots of the different models has been seperated and drawn in the vicinity of each integer degree $l$.  \label{f10} }
\end{figure}

We calculate the eigenfrequencies of the lowest-order p-modes for the objects defined in Table \ref{parametersRadius}, and for $l \in [0,8]$. Figure \ref{f10} presents the corresponding eigenvalues for $n$ up to 12, as a function of the degree $l$, for objects with a radius equal to 1.0, 1.2, 1.4  and 1.6 $R_J$, and a mass equal to 1.0 $M_J$. It appears that the remarks of the previous subsection, which deals with the dependence on the planet mass, also apply to Figure \ref{f10}. Indeed, the frequency spectra of the four planet models appear more and more distinct from one another as we go up in frequency, and as we go up in degree $l$. The frequency range and scale of the low-degree, low-order eigenvalues decrease with the planet radius, when the mass is fixed, whereas the same parameters increase with the planet mass, when the radius is fixed. For instance, numerically, the low-degree, low-order eigenvalues of a 1.6-$R_J$ planet are between 0.07 $mHz$ and 0.91 $mHz$, whereas, in the case of a 1.0-$R_J$ planet, the same modes have eigenfrequencies between 0.1  $mHz$ and 2.6 $mHz$.

\subsubsection{Dependence on the core mass}

Even in the cases of Jupiter and Saturn, the presence and mass of a dense core is still not proven. \citet{gudkova99} have shown that measurements of the pulsation modes of Jupiter could constrain the dimensions of the core. This is likely to be true for exoplanets, if and when their modes are measured. Many extrasolar giant planets appear smaller than the theory would allow \citep{burrows07, guillot06}. This anomaly can be explained by the presence of heavy elements in a dense core, which shrinks the radii of these planets. One famous example is the case of HD149026b, whose measured radius and mass suggest the presence of a core mass in the range 45 - 90 $M_{\oplus}$ \citep{sato05}.

\begin{figure}
\plotone{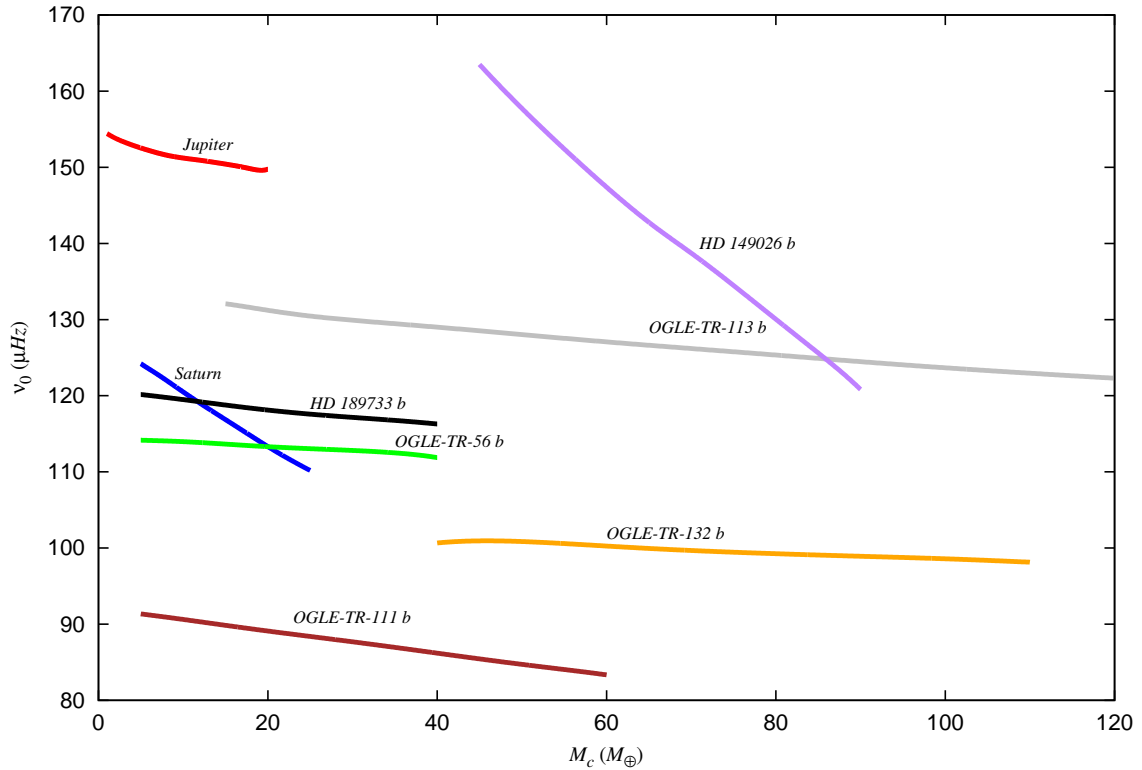}
\caption{The charateristic frequency $\nu_0$ as a function of the core mass, $M_c$, for various exoplanets. The helium mass fraction has been set equal to 0.25 in the envelope. The estimated range of core masses have been taken from various sources;
Jupiter, Saturn: \citet{saumon04}; HD149026b: \citet{sato05}; for all the other planets: \citet{burrows07}. \label{f11} }
\end{figure}

We calculate the characteristic frequency $\nu_0$ as a function of the core mass for a selection of exoplanets for which the presence of a core has been inferred. The results are given in Figure \ref{f11}, which also includes the characteristic frequencies for Jupiter and Saturn, as a reference. The core mass ranges have been taken from \citet{saumon04} for Jupiter and Saturn, \citet{sato05} for HD149026b and \citet{burrows07} for the other planets. It is clear that, in any case, with the radius and the mass of the object fixed, $\nu_0$  is a decreasing function of the core mass. This can be easily explained: the presence of a dense core reduces the sound speed in the center of the planet (see, for example, Figure \ref{f1}) which ultimately increases the integral $\int_{planet} \frac{dr}{c}$ and, thus, diminishes $\nu_0$, which is inversely proportionnal to the latter. However, the sensitivity to the mass of the core is not identical among the selected objects. We can see that Saturn and HD149026b are much more influenced by the core mass than the others. This is due to their small radius and mass, compared to the other selected planets. Indeed, Saturn's radius is $0.83$ $R_J$, its mass is $0.30$ $M_J$,  HD149026b's radius is $0.72$ $R_J$, its mass is $0.36$ $M_J$ whereas all the other planets have radii within the range [1.0,1.23] $R_J$ and masses within the range [0.54,1.30] $M_J$. In this way, for planets with a small radius and mass, the determination of $\nu_0$ through observation can be a powerful tool to investigate the presence of a dense core. For example, for our models of HD149026b, $\nu_0$ loses more than 26\% of its value between a 45-$M_{\oplus}$ core model and a 90-$M_{\oplus}$ core model. Thus, even a rough estimate of the value of $\nu_0$ might give us information on the core of this type of planet.

\begin{figure}
\plotone{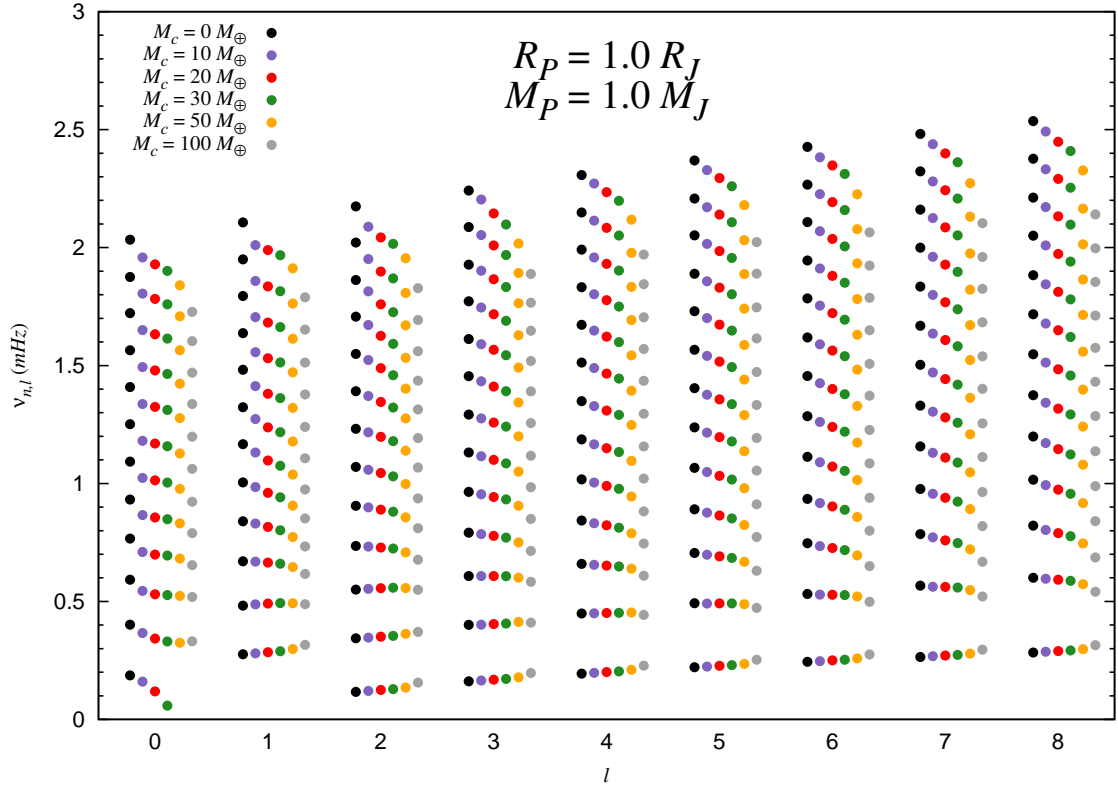}
\caption{Low-order eigenfrequencies of oscillation modes (in $mHz$), as a function of the degree $l$, for exoplanet models with various core masses, with the radius fixed at $R_P = 1.0$ $R_J$ and the mass fixed at $M_P = 1.0$ $M_J$. The helium mass fraction in the envelope is 0.25. For the sake of clarity, the eigenvalue dots of the different models has been seperated and drawn on both sides of each integer degree $l$.  \label{f12} }
\end{figure}

To investigate the dependence of the low-degree, low-order eigenfrequencies on the core mass, we build several exoplanet models with the radius fixed at $R_P = 1.0$ $R_J$,
the mass fixed at $M_P = 1.0$ $M_J$, and the core mass in the range 0-100 $M_{\oplus}$.
Table \ref{parametersCore1} presents the various parameters of these models: the core mass, $M_c$ (in Earth units), the central pressure, $p_c$ (in Mbars), the central density, $\rho_c$ (in cgs units), the specific entropy, $S$  (in $k_B  / baryon$), and the characteristic frequency, $\nu_0$ (in $\mu Hz$).
The lowest-order eigenvalues of modal oscillations are given in Figure \ref{f12}, as a function of the degree $l$, for models with a core mass equal to 0, 10, 20, 30, 50, and 100 $M_{\oplus}$, and with the radius and mass fixed at the Jovian values. When the radius and mass are fixed, the eigenfrequencies are monotonic functions of the core mass, but their direction of variation depends on their degree $l$ and their radial order $n$. For $l \geq 2$, the eigenvalues of the f-modes slowly increase when the core mass increases. At high $n$, for every $l$, it can be seen that the eigenfrequencies are decreasing functions of the core mass, all things being equal. For instance, numerically, between the coreless model and the model with a core mass fixed at 100 $M_{\oplus}$, the eigenvalues decrease by $\sim$$15$\% at high $n$ ($n \in [10,12]$), for every $l \in [0,8]$. Thus, for a given degree $l$, the f-modes and the high-order p-modes, as functions of the core mass, have opposite directions of variation. Consequently, the frequency range for a given $l$ shrinks when the core mass increases. If the low-degree, low-order modes were to be unambiguously identified by its spherical harmonic quantum numbers, the corresponding frequency range may constrain the presence of heavy elements in the deep interior.

We have also constructed several exoplanet models with the mass fixed at $M_P = 1.0$ $M_J$, and the specific entropy fixed at $S=6.67$ $k_B  / baryon$, which is the specific entropy of our coreless model with  $R_P = 1.0$ $R_J$ and $M_P = 1.0$ $M_J$. These models possess a core with a mass in the range  0-100 $M_{\oplus}$ and this set of planet models approximately probes the situation in which, for a given planet mass and a given age, the presence of a dense core shrinks the radius.
Table \ref{parametersCore2} presents various parameters of these models:  the core mass, $M_c$ (in Earth units), the planet radius, $R_P$ (in Jupiter units),  the central pressure, $p_c$ (in Mbars), the central density, $\rho_c$ (in cgs units), and the characteristic frequency, $\nu_0$ (in $\mu Hz$).  For the selected values of planet mass ($M_P = 1.0$ $M_J$) and entropy ($S=6.67$ $k_B  / baryon$), the shrinking of the radius with core mass is visible. For instance, the planet radius decreases by $\sim$$20$ \% when a 100-$M_{\oplus}$ core is added. The direct consequence of the reduction of the radius is the increase of the characteristic frequency $\nu_0$. Qualitatively, the increase of $\nu_0$ is consistent with its dependence on the planet radius, discussed in section \ref{42}.

For the fixed radius and entropy, the eigenvalues of the lowest-order modal oscillations are given in Figure \ref{f12}, as a function of the degree $l$, for models with a core mass equal to 0, 10, 20, 50, and 100 $M_{\oplus}$. When the mass and the entropy of the planet models are fixed, the frequency range of the low-degree, low-order p-modes decreases as the core mass increases. This is again a direct consequence of the shrinking of the planet radius as the core mass goes up, for given mass and entropy. However, as can be seen, the frequency spectra of the models defined by $M_c = 0$ $M_{\oplus}$ and $M_c = 10$ $M_{\oplus}$ are quite similar for $l \in [0,3]$ and $n \in [0,12]$. In particular, for $l=1$ and $2$, the eigenfrequencies of these two models differ by less than 1\% for the frequency range considered. These similarities may be due to two opposite effects. First, we know that the presence of a 10-$M_{\oplus}$ dense core shrinks the model radius by 2\% for the selected values of planet mass and entropy (see Table \ref{parametersCore2}). When the planet mass is fixed, this decrease of the radius causes an increase of the low-degree, low-order eigenvalues, as already discussed in section \ref{42}.
On the other hand, when the planet radius and the planet mass are both fixed, the presence of a dense core implies a decrease of the same eigenvalues. Figure \ref{f12} shows that, for the selected values, for low core mass and low degree $l$, the presence of a dense core compensates for the effect of the radius reduction on the eigenvalues. Nonetheless, it can be seen that, for higher degrees $l$ ($l \geq 4$) and for higher core masses ($M_c \geq 20$ $M_{\oplus}$), the effects of the radius reduction exceed the pure effect of the core.

\subsubsection{Dependence on the metallicity}

If not contained in a dense core, heavy elements can be laced throughout the envelope \citep{guillot05}. In this spirit, we investigate the influence of heavy elements in the envelope itself, regardless of the presence of a core. Though there is no published robust equation of state that properly includes heavy elements beyond helium, we can mimic their presence by using a higher helium mass fraction than $Y = 0.25$ \citep{guillot08, spiegel11}. We assume the excess of helium mass fraction $\Delta Y$ (compared with the default value $Y_0 = 0.25$) is given by the value of the metallicity $Z$:
\begin{equation}
\Delta Y \sim  Z \ . 
\end{equation}
Using the value from \citet{asplund09}, we take  $Z_{\sun} = 0.014$, to be the heavy element mass fraction in the Sun. In this way, an helium fraction of $Y = 0.30$ mimics a metallicity equal to $\sim$$3.6$ $\times$ solar metallicity.

We build three coreless models of exoplanets with a helium mass fraction of 0.25 and 0.30, respectively, with the radius and mass fixed at the Jovian values. The corresponding parameters, in particular the characteristic frequencies $\nu_0$, are given in Table \ref{parametersHe}. As can be seen, a higher helium mass fraction, hence a higher fraction of heavy elements in the envelope, tends to lower the value of $\nu_0$, all things being equal. Figure \ref{f14} portrays the low-degree, low-order eigenfreqdeuencies of the modal oscillations for the models of Table \ref{parametersHe}. It appears that the remarks made concerning Figure \ref{f12}, which deals with the dependence on the core mass, can be also made for Figure \ref{f14}. Such similarities are expected since, in both cases, it is the dependence on the global fraction of heavy elements that is under consideration. As can be seen on  Figure \ref{f14}, for low $l$ and very low $n$, the eigenfrequencies slightly increase with the helium mass fraction $Y$, whereas for higher $l$ and $n$, the eigenvalues unambiguously decrease with $Y$. As previously stated when discussing the presence of a dense core, if the low-degree, low-order modes were to be unambiguously identified, the corresponding frequency spectrum might give us a hint of the presence of heavy elements in the interior.

\subsubsection{Evolution of $\nu_0$ with time}

To investigate the evolution of $\nu_0$ for a given explanet, we build simple evolutionnary models of exoplanets with a mass fixed at 0.5, 1.0, 2.0, 10, and 20 $M_J$. We use the default formalism and modeling tools outlined in \citet{burrows97, burrows01, burrows95}. The planets are in isolation, which means that no stellar irradiation is taken into account. No core has been added. As the specific entropy decreases with time, the planet's radius decreases. Figure \ref{f15} portrays the evolution of $\nu_0$ and the planet radius up to 5 $Gyrs$ for the five fixed planet masses considered. The large early radii of the models result in small values for the characteristic frequencies, compared to their  final values. Numerically, $\nu_0(0)$ is between 16\% and 22\% of the final $\nu_0$ values, for the five models considered here. As the radius stabilizes, so does $\nu_0$. We fit the curves with straight lines in the region [2,5] Gyrs, and we calculate the corresponding derivatives. We find that the derivative increases with the planet mass, from $\sim$$3.1$ $\mu Hz \ Gyrs^{-1}$ for $M_P = 0.5$ $M_J$, to  $\sim$$8.2$ $\mu Hz \ Gyrs^{-1}$ for $M_P = 20$ $M_J$.

\section{Conclusion}

We have calculated the eigenfrequencies and eigenfunctions of the pulsational modes of planets for a broad range of giant exoplanet models. In particular, we have investigated the dependence of the characteristic frequency $\nu_0$,  the eigenfrequencies of low degree f-modes, and the eigenfrequencies of low degree p-modes on several parameters: the planet mass, the planet radius, the core mass, and the helium fraction in the envelope. We provide the corresponding eigenvalues for a degree $l$ up to 8 and a radial order $n$ up to 12. We also present values of $\nu_0$ for 174 known giant exoplanets, and highlight the strong dependence on the radius, around any value of the planet mass. 

For Jupiter and Saturn, we find that our results are consistent with both observationnal data \citep{gaulme11} and previous theoretical work \citep{provost93, mosser94, gudkova99}. In the specific case of Jupiter, we presented $\nu_0$ and the low-degree, low-order eigenfrequencies of acoustic modes as a function of the core mass. We conclude that, for nonradial oscillations ($l \geq 1$), the sensitivity of the pulsation frequencies to the core mass is important both for the f-modes and for the high-order p-modes. However, for specific intermediate values of radial order $n$, this sensitivity is minimal over the whole range of core masses considered.

Focusing on giant exoplanets, we find that the dependence of the characteristic frequency on the core mass is more important for small radii and masses. As an example, the characteristic frequency of HD149026b, with measured radius and mass of 0.72 $R_J$ and 0.36 $M_J$, varies by more than 26\% across the range of core masses considered. We quantify the influence of the core mass on exoplanet models with arbitrary fixed mass and radius. For $l \in [0,8]$ and for frequencies up to 2.6 $mHz$, we find that eigenfrequencies shrink as the core mass increases, which is consistent with previous work on Jupiter. A big core ($M_c = 100$ $M_{\oplus}$) induces a reduction in $\nu_{n,l}$ of $\sim$$15$\% for $n \geq 10$ compared to coreless values. We also develop an approach to quantify the influence  on the eigenfrequency spectrum of a high heavy-element fraction in the planet envelope. We find that, quantitatively, the presence of heavy element in the envelope affects the eigenvalue distribution in ways similar to the presence of a dense core.

We have also quantified the influence of mass and radius on the modal oscillations of giant exoplanets. For the selected values of $l$ and $n$, we find that the pulsation eigenfrequencies depend strongly on both parameters, especially at high frequency. This dependence can be measured through $\nu_0$. For the mass range $0.5 \leq M_P \leq 15$ $M_J$,  and fixing the planet radius to its Jovian value, we find that $\nu_0 \sim 164.0 \times \left(M_P/M_J\right)^{0.48} \mu Hz$. For the radius range from 0.9 to 2.0 $R_J$, and fixing the planet's mass to its Jovian value, we find that $\nu_0 \sim 164.0 \times \left(R_P/R_J\right)^{-2.09} \mu Hz$. These variations of $\nu_0$ directly affect the high-frequency spectrum of modal oscillations.

\begin{acknowledgments}

We thank Dave Spiegel and Sudhir Raskutti for helpful discussions. The 
authors would also like to acknowledge support in part under HST grants 
HST-GO-12181.04-A, HST-GO-12314.03-A, and HST-GO-12550.02, and JPL/Spitzer 
Agreements 1417122, 1348668, 1371432, 1377197, and 1439064.

\end{acknowledgments}

\clearpage

\begin{appendix}

\subsection{NONRADIAL OSCILLATION EIGENVALUE PROBLEM}

For adiabatic, nonradial oscillations of non rotating spherically symmetrical planetary models, the governing differential equations are \citep[eqs. 14.2 - 14.4]{unno89}:
\begin{equation}
\frac{1}{r^2} \frac{d}{dr}\left(r^2 \xi _r\right) - \frac{g}{c^2}\xi _r + \left(1 - \frac{L_l^2}{\sigma^2} \right)\frac{p'}{\rho c^2} = \frac{l(l+1)}{\sigma^2 r^2}\Phi',   \label{U1}
\end{equation}
\begin{equation}
\frac{1}{\rho} \frac{dp'}{dr} +  \frac{g}{\rho c^2}p' + \left(N^2 - \sigma^2\right)\xi _r =  - \frac{d\Phi'}{dr},                                                                                                                               \label{U2}
\end{equation}
and
\begin{equation}
\frac{1}{r^2} \frac{d}{dr}\left(r^2 \frac{d\Phi'}{dr}\right) - \frac{l(l+1)}{ r^2}\Phi' = 4 \pi G \rho \left(\frac{p'}{\rho c^2} + \frac{N^2}{g} \xi_r\right),          \label{U3}
\end{equation}
where $c =(\Gamma p_0 /  \rho_0)^{1/2}$ is the sound speed and the Lamb frequency, $L_l$, is:
\begin{equation}
L_l^2 = \frac{l(l+1)c^2}{r^2}    \label{Lamb}
\end{equation}
and the Brunt-V\"ais\"al\"a frequency, N, is:
\begin{equation}
N^2 = g \left(\frac{1}{\Gamma}\frac{d\ln p_0}{dr} - \frac{d\ln\rho_0}{dr}\right).  \label{Brunt}
\end{equation}

$\Gamma = (d\ln p_0/d\ln\rho_0)_{ad}$ is the adiabatic exponent, G is the gravitation constant, $g = GM_r/r^2$ is the
gravitational acceleration, and $r$ is the radius. Primed variables refer to the
Eulerian perturbation at a given position; zero subscripts refer to the equilibrium
value.

Equations \eqref{U1} - \eqref{U3} are the full fourth-order set of differential equations. There
are four corresponding boundary conditions \citep[eqs. 14.8-14.11]{unno89}.
At $r=0$,
\begin{equation}
\xi_r - \frac{l}{\sigma^2 r} \left(\frac{p'}{\rho} + \Phi' \right) = 0  \label{B1}
\end{equation}
and
\begin{equation}
\frac{d\Phi'}{dr} - \frac{l\Phi'}{r} = 0 \ .  \label{B2}
\end{equation}

At $r=R$, where $R$ is the radius of the planet,
\begin{equation}
\frac{d\Phi'}{dr} + \frac{l(l+1)}{r} \Phi' = 0  \label{B3}
\end{equation}
and
\begin{equation}
\delta p = 0.  \label{B4}
\end{equation}
Here, $\delta$  is the Lagrangian perturbation for a given fluid element. Equation \eqref{B4}  has
various limiting forms, depending on the physical conditions at $ r = R$. For the
case in which the density and pressure vanish at the surface, \eqref{B4} can be written
\citep[ eq. 14.12]{unno89}:
\begin{equation}
\xi_r - \frac{p'}{g\rho} = 0 \ .  \label{B4bis}
\end{equation}
This condition is valid whenever
\begin{equation}
- \frac{d\ln p_0}{d\ln r} = \frac{r}{H} \gg 1  \label{Condition1}
\end{equation}
at $r = R$. Thus, we need to estimate the pressure scale height, $H$, for every input model of the planet. As an example, Saturn's value is about 40 km at 1 bar \citep{marley90}, making this condition for a free boundary appropriate. The differential
equations, boundary conditions, and a normalization condition at $r = R$, $ \xi_r / r = 1$,
comprise the eigenvalue problem.

%%%%%%%%%%%%%%

\subsection{EIGENVALUE CALCULATION: THE DIMENSIONLESS PROBLEM}

The eigenvalue differential equations (Eqs.\ref{U1} - \ref{U3}) may be recast as four first-order
differential equations with four dimensionless variables \citep{unno89}.
The variables are:
\begin{eqnarray}
y_1 & = & \frac{\xi_r}{r},  \nonumber \\
y_2 & = & \frac{1}{gr} \left(\frac{p'}{\rho} + \Phi'\right),  \nonumber \\
y_3 & = & \frac{1}{gr} \Phi',  \nonumber \\
y_4 & = & \frac{1}{g} \frac{d\Phi'}{dr}.  \label{V4}
\end{eqnarray}
The dimensionless variable
\begin{equation}
x = \frac{r}{R}
\end{equation}
is used in place of $r$. The resulting four equations are as follows:
\begin{equation}
x \frac{dy_1}{dx} = \left(V_g - 3 \right) y_1 + \left\lbrack\frac{l(l+1)}{c_1\omega^2} - V_g\right\rbrack y_2+ V_g  y_3,  \label{UD1}
\end{equation}
\begin{equation}
x \frac{dy_2}{dx} = \left(c_1\omega^2 - A^* \right) y_1 + \left( A^* - U +1 \right) y_2+ A^*  y_3, \label{UD2}
\end{equation}
\begin{equation}
x \frac{dy_3}{dx} = \left(1-U \right) y_3 +  y_4, \label{UD3}
\end{equation}
and

\begin{equation}
x \frac{dy_4}{dx} = U A^*  y_1 + UV_gy_2+ \left\lbrack l(l+1) -UV_g\right\rbrack y_3+ U  y_4. \label{UD4}
\end{equation}

The dimensionless quantities are
\begin{equation}
V_g =  \frac{d \ln M_r}{d \ln r} = \frac{gr}{c^2} \label{DQ1}
\end{equation}
\begin{equation}
U = - \frac{1}{\Gamma} \frac{d \ln p}{d \ln r} = \frac{4 \pi \rho r^3}{M_r} \label{DQ2}
\end{equation}
\begin{equation}
c_1 = \left(r/R\right)/\left(M_r/M\right)  \label{DQ3}
\end{equation}
\begin{equation}
\omega^2 = \frac{\sigma^2 R^3}{GM} = \frac{4 \pi^2\nu^2 R^3}{GM}  \label{DQ41}
\end{equation}
and
\begin{equation}
A^* = -rA = rg^{-1}N^2.  \label{DQ5}
\end{equation}

The dimensionless boundary conditions are
\begin{eqnarray}
\frac{c_1 \omega ^2}{l} y_1 - y_2 & = 0  \text{    at } &  r = 0 \label{BC1} \\
l y_3 - y_4 & = 0  \text{    at } &  r = 0 \label{BC2} \\
(l + 1)y_3 + y_4 & = 0  \text{    at } &  r = R \label{BC3}\\
y_1 - y_2 + y_3 & = 0  \text{    at } &  r = R. \label{BC4}
\end{eqnarray}

\clearpage
%%%%%%%%%%%%%%%%%%%%%%%%%%%%%%%%%%%%%%%%%%%%%%%%%%%

\subsection{NUMERICAL IMPLEMENTATION OF THE OSCILLATION SOLUTIONS}

Using the previous equations and boundary conditions, we build two linearly independent solutions that satisfy the appropriate boundary conditions at the
center and two linearly independent solutions that satisfy the appropriate boundary conditions at the surface. These solutions are carried out using a fifth-order Runge-Kutta method with adjusted stepsize to ensure accuracy.
The solution vectors  $\mathbf{y} = (y_i)$ are, respectively: $\mathbf{y}^{C,1}(x)$, $\mathbf{y}^{C,2}(x)$, $\mathbf{y}^{S,1}(x)$, and $\mathbf{y}^{S,2}(x)$, where the superscripts $C$ and $S$ denote a solution integrated from the center 
and the surface, respectively.
A continuous match of the interior and exterior solutions at an arbitrary fitting point $x_f$ 
requires the existence of non-zero constants $K_i^{C,1}$, $K_i^{C,2}$, $K_i^{S,1}$, and  $K_i^{S,1}$, $i \in \{1,2,3,4\}$ such that \citep{dalsgaard03}:
\begin{equation}
K_i^{C,1} y_i^{C,1}(x_f) + K_i^{C,2} y_i^{C,2}(x_f) = K_i^{S,1} y_i^{S,1}(x_f) + K_i^{S,2} y_i^{S,2}(x_f), \label{continuous}
\end{equation}
For all $i \in  \{1,2,3,4\}$. This set of equations has a solution only if the determinant
\begin{equation}
\Delta_f(\omega^2) = \left| \mathbf{y}^{C,1}(x_f)  \quad \mathbf{y}^{C,2}(x_f)  \quad  \mathbf{y}^{S,1}(x_f) \quad  \mathbf{y}^{S,2}(x_f)  \right|   \label{determinant}
\end{equation}
vanishes. Hence, the eigenfrequencies are determined as the zeros of $\Delta_f(\omega^2)$.
This determinant has the advantage of behaving smoothly over the whole frequency spectrum and allows us to use a simple bisection method to find all the roots of $\Delta_f$, while scanning a given interval of frequency.

The roots of $\Delta_f$ are supposed to be independent both of the choice of the fitting point and of the initial values of  $\mathbf{y}^{C,1}$, $\mathbf{y}^{C,2}$, $\mathbf{y}^{S,1}$ and $\mathbf{y}^{S,2}$.
However, it is possible to control the amplitude of the determinant by using the regular solutions near the center and the surface, given in Unno et al. 1989 and characterized by:
\begin{eqnarray}
y_1 & \sim & x^{l-2}  \qquad \text{for }  r \sim 0,  \\
y_1 &  \sim &  x^{-l}  \qquad \text{for }  r \sim R.
\end{eqnarray}
Thus, we define the following initial values for the center:
\begin{eqnarray}
y_1 ^{C,1} & = & y_1 ^{C,2}  =  x^{l-2}, \\
y_3^{C,1} & = & f_1 \cdot y_1 ^{C,1},\\
y_3^{C,2} & = & f_2 \cdot y_1 ^{C,1},
\end{eqnarray}
and for the surface:
\begin{eqnarray}
y_1 ^{S,1} &  = & y_1 ^{S,2}  =  x^{l-2}, \\
y_3^{S,1} & = & g_1 \cdot y_1 ^{S,1},\\
y_3^{S,2} & = & g_2 \cdot y_1 ^{S,1},
\end{eqnarray}
where $f_1$, $f_2$, $g_1$ and $g_2$ are arbitrary coefficients such that $f_1 \neq f_2$ and $g_1 \neq g_2$.
Then, for any particular frequency (actually $\omega^2$), the values of $y_i ^{C,j}$ and $y_i ^{S,j}$ for $i=2,4$ and $j=1,2$ are fixed by the boundary conditions \eqref{BC1} - \eqref{BC4}.

\subsection{NUMERICAL IMPLEMENTATION OF THE PLANET INTERIOR PROFILES}

Using the classical equations of the hydrostatic equilibrium, the interior profiles are derived using a fifth-order Runge-Kutta method with adjusted stepsize to ensure accuracy. As a consequence, the number of points of the radial grid is not constant, but is about 1500 points. The density, pressure, sound speed, and gravitational acceleration are obtained by linear interpolation.
Three layers are considered for the planet interiors: an adiabatic atmosphere, a hydrogen-helium envelope, and an olivine core. If a core mass is specified, the adaptive stepsize permits the code to carry out the profile solution from the center to the exact radius interior to which the specified core mass is reached. At this point, pressure continuity is ensured, whereas density, temperature, and sound speed are recalculated using the equation of state of the hydrogen-helium envelope.
The transition between the atmosphere and the envelope has been smoothed through linear interpolation to ensure the continuity of density, pressure, and sound speed.

\end{appendix}
\clearpage

%%%%%%%%%%%%%%%%%%%%%%%%%%%

\clearpage

%%%%%%%%%%%%%%%%%%%%%%%%%%%%%%%%%%%%%%%%%%%%%%%%%%%%%%%%%%%%%%%%%%%%%%%%%%%%%%%%%%%%%%%%%%%%%%%

\begin{deluxetable}{ccccc}
\tablewidth{0pt}
\tablecaption{Eigenfrequencies of  p-modes of degree $0$ and $1$ for a polytrope of index 3, in $\mu Hz$.The radius is  $R_P  =  6.9599 \times 10^{10} \ cm$ and the mass is $M_P = 1.989 \times 10^{33} \ g$.   \label{pmodes1}}
\tablehead{
$l$ & \multicolumn{2}{c}{\underline{0}} &  \multicolumn{2}{c}{\underline{1}} \\
\colhead{$n$}           & \colhead{ This work}      &
\colhead{CDM$^*$}          & \colhead{ This work}  &
\colhead{CDM$^*$}}
\startdata
1 & 303.7754 & 303.7755 & 337.2152 & 337.2152\\
2 & 411.5268 & 411.5269 & 463.5716 & 463.5718\\
3 & 532.9179 & 532.9181 & 590.0692 & 590.0694 \\
4 & 658.1677 & 658.1679 & 716.6286 & 716.6289 \\
5 & 784.3664 & 784.3667 & 843.1062 & 843.1066  \\
6 & 910.7238 & 910.7242 & 969.4328 & 969.4331\\
7 & 1036.9925 & 1036.9929 & 1095.5829 & 1095.5832\\
8 & 1163.0917 & 1163.0922 & 1221.5526 & 1221.5530 \\
9 & 1289.0005 & 1289.0010 & 1347.3477 & 1347.3484\\
10 & 1414.7209 & 1414.7214 & 1472.9793 & 1472.9799\\
11 & 1540.2639 & 1540.2645 & 1598.4587 & 1598.4594\\
12 & 1665.6437 & 1665.6443 & 1723.7983 & 1723.7990 \\
13 & 1790.8749 & 1790.8756 & 1849.0098 & 1849.0105\\
14 & 1915.9720 & 1915.9728 & 1974.1042 & 1974.1051\\
15 & 2040.9483 & 2040.9489 & 2099.0918 & 2099.0928\\
16 & 2165.8149 & 2165.8160 & 2223.9823 & 2223.9831\\
17 & 2290.5839 & 2290.5848 & 2348.7834 & 2348.7844\\
18 & 2415.2641 & 2415.2651 & 2473.5035 & 2473.5043\\
19 & 2539.8643 & 2539.8653 & 2598.1491 & 2598.1500\\
20 & 2664.3924 & 2664.3933 & 2722.7266 & 2722.7276\\
21 & 2788.8549 & 2788.8558 & 2847.2419 & 2847.2430\\
22 & 2913.2577 & 2913.2588 & 2971.6997 & 2971.7011\\
23 & 3037.6068 & 3037.6078 & 3096.1059 & 3096.1067\\
24 & 3161.9068 & 3161.9076 & 3220.4627 & 3220.4639\\
25 & 3286.1614 & 3286.1626 & 3344.7754 & 3344.7767
\enddata
\tablerefs{
$^*$\citet[Table 2]{dalsgaard94}.}
\end{deluxetable}

\begin{deluxetable}{ccccc}
\tablewidth{0pt}
\tablecaption{Eigenfrequencies of  p-modes of degree $2$ and $3$ for a polytrope of index 3, in $\mu Hz$. The radius is  $R_P  =  6.9599 \times 10^{10} \ cm$ and the mass is $M_P = 1.989 \times 10^{33} \ g$. \label{pmodes2}}
\tablehead{
$l$ & \multicolumn{2}{c}{\underline{2}} &  \multicolumn{2}{c}{\underline{3}} \\
\colhead{$n$}           & \colhead{ This work}      &
\colhead{CDM$^*$}          & \colhead{ This work}  &
\colhead{CDM$^*$}}
\startdata
1 & 390.1222 & 390.1223 & 428.8390 & 428.8391\\
2 & 516.1991 & 516.1992 & 558.2978 & 558.2981\\
3 & 643.0675 & 643.0677 & 686.8728 & 686.8732\\
4 & 769.7799 & 769.7802 & 814.7025 & 814.7027\\
5 & 896.2907 & 896.2910 & 942.0263 & 942.0267\\
6 & 1022.6131 & 1022.6135 & 1068.9862  & 1068.9866\\
7 & 1148.7618 & 1148.7622 & 1195.6634 & 1195.6639\\
8 & 1274.7484 & 1274.7491 & 1322.1083  & 1322.1090\\
9 & 1400.5845 & 1400.5849 & 1448.3544 & 1448.3552\\
10 & 1526.2791 & 1526.2795 & 1574.4258  & 1574.4265\\
11 & 1651.8419 & 1651.8424 & 1700.3401 & 1700.3411\\
12 & 1777.2824 & 1777.2825 & 1826.1125 & 1826.1136\\
13 & 1902.6072 & 1902.6082 & 1951.7557& 1951.7561\\
14 & 2027.8269 & 2027.8277 & 2077.2784 & 2077.2793\\
15 & 2152.9475 & 2152.9485 & 2202.6912 & 2202.6923\\
16 & 2277.9762 & 2277.9776 & 2328.0025  & 2328.0032\\
17 & 2402.9209 & 2402.9216 & 2453.2187 & 2453.2197\\
18 & 2527.7854 & 2527.7867 & 2578.3473 & 2578.3482\\
19 & 2652.5775 & 2652.5785 & 2703.3935  & 2703.3949\\
20 & 2777.3017 & 2777.3023  & 2828.3646  & 2828.3655\\
21 & 2901.9614 & 2901.9629 & 2953.2632 & 2953.2651\\
22 & 3026.5636 & 3026.5648 & 3078.0968 & 3078.0983\\
23 & 3151.1104 & 3151.1121 & 3202.8689 & 3202.8695\\
24 & 3275.6070 & 3275.6085 & 3327.5814 & 3327.5828\\
25 & 3400.0566 & 3400.0577 & 3452.2396 & 3452.2418
\enddata
\tablerefs{
$^*$ \citet[Table 2]{dalsgaard94}.}
\end{deluxetable}

\begin{deluxetable}{ccccc}
\tablewidth{0pt}
\tablecaption{Eigenfrequencies of  g-modes for a polytrope of index 3, in $\mu Hz$. The radius is  $R_P  =  6.9599 \times 10^{10} \ cm$ and the mass is $M_P = 1.989 \times 10^{33} \ g$. \label{gmodes}}
\tablehead{
$l$ & \multicolumn{2}{c}{\underline{2}} &  \multicolumn{2}{c}{\underline{3}} \\
\colhead{$n$}           & \colhead{ This work}      &
\colhead{CDM$^*$}          & \colhead{ This work}  &
\colhead{CDM$^*$}}
\startdata
-20 & 31.1798 & 31.1797 & 43.0341 & 43.0356\\
-19 & 32.6437 & 32.6439 & 45.0003 & 45.0009\\
-18 & 34.2535 & 34.2535 & 47.1556 & 47.1552\\
-17 & 36.0311 & 36.0313 & 49.5267 & 49.5272\\
-16 & 38.0054 & 38.0053 & 52.1528 & 52.1519\\
-15 & 40.2096 & 40.2099 & 55.0737 & 55.0718\\
-14 & 42.6886 & 42.6884 & 58.3407 & 58.3400\\
-13 & 45.4951 & 45.4953 & 62.0230 & 62.0230\\
-12 & 48.7014 & 48.7010 & 66.2057 & 66.2053\\
-11 & 52.3970 & 52.3972 & 70.9964 & 70.9960\\
-10 & 56.7061 & 56.7065 & 76.5390 & 76.5389\\
-9 & 61.7959 & 61.7959 & 83.0266 & 83.0267\\
-8 & 67.8987 & 67.8991 & 90.7245 & 90.7243\\
-7 & 75.3534 & 75.3535 & 100.0041 & 100.0058\\
-6 & 84.6648 & 84.6649 & 111.4182 & 111.4181\\
-5 & 96.6263 & 96.6264 & 125.7916 & 125.7915\\
-4 & 112.5543 & 112.5543 & 144.4538 & 144.4530\\
-3 & 134.7963 & 134.7963 & 169.6560 & 169.6563\\
-2 & 167.9278 & 167.9279 & 205.5287 & 205.5286\\
-1 & 221.3677 & 221.3677 & 259.7578 & 259.7578
\enddata
\tablerefs{
$^*$ \citet[Table 4]{dalsgaard94}.}
\end{deluxetable}

%%%%%%%%%%%%%%%%%%%%%%%%%%%%%%%%%%%%%%%%%%%%%%%%%%%%%%%%%%%%%%%%%%%%%%%%%%%%%%%%%%%%%%%%%%%%%%%

\begin{deluxetable}{c c c c c c}
\tablewidth{0pt}
\tablecaption{Parameters of the models of Jupiter (J) and Saturn (S). \label{parameters}}
\tablehead{
\colhead{Model}  & \colhead{$Y$}      &
\colhead{$M_{core}$ ($M_\oplus$)}         &
\colhead{$p_c$ ($Mbars$)} & \colhead{$\rho _c$ ($g \ cm^{-3}$)}
&   \colhead{ $\nu _0$ ($\mu Hz$)}}
\startdata
J1 &  0.25 &  $0.0$ & $39.2$ & $4.14$ &  $157.4$ \\
J2 &  0.30 &  $0.0$ & $41.7$ & $4.25$ &  $153.0$ \\
J3 &  0.25 &  $5.0$ & $71.6$ & $19.4$ &  $152.4$ \\
J4 &  0.25 &  $10.0$ & $96.1$ & $21.8$ & $151.2$\\
\hline
S1 & 0.25 & $13.0$ & $44.2$ & $16.1$ &  $118.6$\\
S2 & 0.25 & $18.1$ & $59.8$ & $18.1$  & $115.0$\\
S3 & 0.30 & $18.1$ & $60.7$ & $18.2$  & $112.2$
\enddata
\end{deluxetable}

%%%%%%%%%%%%%%%%%%%%%%%%%%%%%%%%%%%%%%%%%%%%%%%%%%%%%%%%%%%%%%%%%%%%%%%%%%%%%%%%%%%%%%%%%%%%%%%

\begin{deluxetable}{ccccccc}
\tablewidth{0pt}
\tablecaption{Periods of  p-modes for the J4 model (in min).   \label{pmodesJupiter}}
\tablehead{
\colhead{$l$} & \colhead{0} &  \colhead{1} &   \colhead{2} &   \colhead{3} & \colhead{4} & \colhead{5}  \\
\colhead{$n$}   & \multicolumn{6}{c}{ }}
\startdata
0 &	-	&	-	&	138.38 &	101.18  &	84.49	  &	74.46 \\
1 &    104.31 &	59.44	&	48.07	  &	41.52	  &	37.11	  &	33.93 \\
2 &    45.53    &	34.19	&	30.13	 &	27.41	  &	25.42	  &	23.89 \\
3 &    30.61	&	24.93	&	22.74	&	21.21	  &	20.00	  &	19.00 \\
4  &   23.50	&	20.08	&	18.55	&	17.46	  &	16.60	  &	15.88 \\
5 &   19.24 	&	16.92	&	15.75	&	14.94	  &	14.26	 &	13.69 
\enddata
\end{deluxetable}

\clearpage

\begin{deluxetable}{cccc}
\tablewidth{0pt}
\tablecaption{Characteristic frequency $\nu_0$ (in $\mu Hz$) for a set of coreless models using the estimated radius and mass (here in Jupiter units) of detected giant exoplanets. \label{nu0exoplanets1}}
\tablehead{
\colhead{Planet}  & \colhead{$R_{planet}$ ($R_J$)}      &
\colhead{$M_{planet}$ ($M_J$)} 
&   \colhead{ $\nu _0$ ($\mu Hz$)} 
}
\startdata
Kepler-9c   & 0.823 & 0.171 & 94.40 \\ 
HAT-P-18b   & 0.995 & 0.197 & 65.01 \\ 
HAT-P-12b   & 0.959 & 0.211 & 73.14 \\ 
Kepler-34b   & 0.764 & 0.220 & 149.0 \\ 
WASP-29b   & 0.792 & 0.244 & 141.9 \\ 
Kepler-9b   & 0.842 & 0.252 & 114.1 \\ 
HAT-P-38b   & 0.825 & 0.267 & 129.8 \\ 
WASP-39b   & 1.270 & 0.280 & 47.87 \\ 
HAT-P-19b   & 1.132 & 0.292 & 61.16 \\ 
WASP-20b   & 0.900 & 0.300 & 105.4 \\ 
WASP-21b   & 1.210 & 0.300 & 54.29 \\ 
WASP-69b   & 1.000 & 0.300 & 80.85 \\ 
HD-149026b   & 0.718 & 0.356 & 229.4 \\ 
WASP-49b   & 1.115 & 0.378 & 72.11 \\ 
WASP-63b   & 1.430 & 0.380 & 44.52 \\ 
WASP-67b   & 1.400 & 0.420 & 48.45 \\ 
Kepler-12b   & 1.695 & 0.431 & 35.07 \\ 
Kepler-7b   & 1.614 & 0.433 & 38.18 \\ 
WASP-11b   & 1.045 & 0.460 & 92.36 \\ 
CoRoT-5b   & 1.388 & 0.467 & 51.76 \\ 
WASP-31b   & 1.537 & 0.478 & 43.59 \\ 
WASP-13b   & 1.365 & 0.485 & 54.39 \\ 
WASP-17b   & 1.991 & 0.486 & 33.51 \\ 
WASP-60b   & 0.900 & 0.500 & 143.5 \\ 
WASP-42b   & 1.080 & 0.500 & 89.56 \\ 
WASP-52b   & 1.300 & 0.500 & 60.66 \\ 
WASP-6b   & 1.224 & 0.503 & 68.68 \\ 
KOI-254b   & 0.960 & 0.505 & 120.4 \\ 
HAT-P-1b   & 1.217 & 0.524 & 70.94 \\ 
HAT-P-17b   & 1.010 & 0.534 & 108.5 \\ 
CoRoT-16b   & 1.170 & 0.535 & 77.76 \\ 
OGLE-TR-111b   & 1.077 & 0.540 & 93.89 \\ 
WASP-15b   & 1.428 & 0.542 & 52.87 \\ 
HAT-P-25b   & 1.190 & 0.567 & 77.34 \\ 
WASP-62b   & 1.390 & 0.570 & 56.96 \\ 
WASP-55b   & 1.300 & 0.570 & 64.76 \\ 
WASP-25b   & 1.260 & 0.580 & 69.56 \\ 
WASP-22b   & 1.158 & 0.588 & 83.46 \\ 
WASP-34b   & 1.220 & 0.590 & 74.97 \\ 
WASP-56b   & 1.200 & 0.600 & 78.27 \\ 
WASP-54b   & 1.400 & 0.600 & 57.66 \\ 
Kepler-8b   & 1.419 & 0.603 & 56.37 \\ 
XO-2b   & 0.973 & 0.620 & 130.3 \\ 
HAT-P-28b   & 1.212 & 0.626 & 78.36 \\ 
Kepler-15b   & 0.960 & 0.660 & 139.9 \\ 
HAT-P-27b   & 1.055 & 0.660 & 109.8 \\ 
Kepler-6b   & 1.323 & 0.669 & 67.80 \\ 
HAT-P-9b   & 1.400 & 0.670 & 60.92 \\ 
OGLE-TR-10b   & 1.720 & 0.680 & 42.53 \\ 
HAT-P-4b   & 1.270 & 0.680 & 74.19 \\ 
HAT-P-24b   & 1.242 & 0.685 & 77.99 \\ 
WASP-59b   & 0.900 & 0.700 & 181.6 \\ 
HAT-P-30b   & 1.340 & 0.711 & 68.21 \\ 
HD-209458b   & 1.380 & 0.714 & 64.62 \\ 
WASP-35b   & 1.320 & 0.720 & 70.67 \\ 
CoRoT-4b   & 1.190 & 0.720 & 87.62 \\ 
OGLE-TR-211b   & 1.260 & 0.750 & 79.26 \\ 
TrES-1   & 1.099 & 0.761 & 107.7 \\ 
HAT-P-33b   & 1.827 & 0.763 & 40.7 \\ 
HAT-P-29b   & 1.107 & 0.778 & 107.1 \\ 
WASP-68b   & 0.900 & 0.800 & 196.7 \\ 
WASP-57b   & 1.100 & 0.800 & 110.3 \\ 
CoRoT-9b   & 1.050 & 0.840 & 126.6 \\ 
WASP-2b   & 1.079 & 0.847 & 119.2 \\ 
HAT-P-13b   & 1.280 & 0.850 & 81.77 \\ 
WASP-16b   & 1.008 & 0.855 & 141.4 \\ 
WASP-1b   & 1.484 & 0.860 & 61.93 \\ 
KOI-202b   & 1.020 & 0.880 & 139.4 \\ 
WASP-23b   & 0.962 & 0.884 & 163.6 \\ 
WASP-44b   & 1.140 & 0.889 & 107.5 \\ 
WASP-79b   & 1.700 & 0.890 & 49.60 \\ 
XO-1b   & 1.184 & 0.900 & 99.52 \\ 
WASP-28b   & 1.120 & 0.910 & 113.3 \\
TrES-4   & 1.706 & 0.917 & 50.07
\enddata
\tablecomments{The helium mass fraction has been fixed at 0.25 in the entire envelope. No core has been added. Estimated radii and masses have been taken from \emph{http://www.exoplanet.eu}}
\end{deluxetable}

\clearpage

\begin{deluxetable}{cccc}
\tablewidth{0pt}
\tablenum{6B}
\tablecaption{Characteristic frequency $\nu_0$ (in $\mu Hz$) for a set of coreless models using the estimated radius and mass (here in Jupiter units) of detected giant exoplanets}
\tablehead{
\colhead{Planet}  & \colhead{$R_{planet}$ ($R_J$)}      &
\colhead{$M_{planet}$ ($M_J$)} 
&   \colhead{ $\nu _0$ ($\mu Hz$)} 
}
\startdata
CoRoT-12b   & 1.440 & 0.917 & 67.65 \\
WASP-41b   & 1.210 & 0.920 & 95.95 \\ 
HAT-P-32b   & 2.037 & 0.941 & 44.39 \\ 
WASP-7b   & 1.330 & 0.960 & 80.61 \\ 
WASP-48b   & 1.670 & 0.980 & 53.73 \\ 
WASP-45b   & 1.160 & 1.007 & 110.4 \\ 
KOI-204b   & 1.240 & 1.020 & 95.95 \\ 
WASP-26b   & 1.281 & 1.028 & 90.00 \\ 
CoRoT-1b   & 1.490 & 1.030 & 67.42 \\ 
WASP-24b   & 1.104 & 1.032 & 125.3 \\ 
HAT-P-35b   & 1.332 & 1.054 & 84.32 \\ 
HAT-P-6b   & 1.330 & 1.057 & 84.70 \\ 
OGLE-TR-182b   & 1.470 & 1.060 & 70.14 \\ 
HAT-P-5b   & 1.252 & 1.060 & 95.89 \\ 
Qatar-1b   & 1.164 & 1.090 & 114.2 \\ 
WASP-58b   & 1.300 & 1.100 & 90.49 \\ 
CoRoT-19b   & 1.450 & 1.110 & 73.68 \\ 
WASP-4b   & 1.363 & 1.121 & 83.28 \\ 
HD-189733b   & 1.138 & 1.138 & 122.9 \\ 
WASP-47b   & 1.150 & 1.140 & 120.2 \\ 
WASP-78b   & 1.750 & 1.160 & 54.10 \\ 
WASP-19b   & 1.386 & 1.168 & 82.39 \\ 
HAT-P-37b   & 1.178 & 1.169 & 115.3 \\ 
OGLE-TR-132b   & 1.230 & 1.170 & 104.8 \\ 
OGLE-TR-113b   & 1.110 & 1.240 & 136.2 \\ 
TrES-2   & 1.169 & 1.253 & 121.6 \\ 
OGLE-TR-56b   & 1.200 & 1.300 & 116.9 \\ 
CoRoT-13b   & 0.885 & 1.308 & 562.3 \\ 
HAT-P-8b   & 1.500 & 1.340 & 76.46 \\ 
WASP-12b   & 1.736 & 1.404 & 60.61 \\ 
WASP-50b   & 1.153 & 1.468 & 136.3 \\ 
KELT-2Ab   & 1.306 & 1.486 & 104.9 \\ 
WASP-65b   & 1.300 & 1.600 & 110.1 \\ 
WASP-5b   & 1.171 & 1.637 & 139.2 \\ 
WASP-37b   & 1.136 & 1.696 & 151.9 \\ 
TrES-5   & 1.209 & 1.778 & 135.3 \\ 
HAT-P-7b   & 1.421 & 1.800 & 98.83 \\ 
HAT-P-36b   & 1.264 & 1.832 & 125.2 \\ 
HATS-1b   & 1.302 & 1.855 & 118.8 \\ 
TrES-3   & 1.305 & 1.910 & 120.1 \\ 
HAT-P-15b   & 1.072 & 1.946 & 187.1 \\ 
WASP-43b   & 1.036 & 2.034 & 207.8 \\ 
WASP-61b   & 1.240 & 2.060 & 138.6 \\ 
WASP-3b   & 1.454 & 2.060 & 101.8 \\ 
HAT-P-23b   & 1.368 & 2.090 & 115.0 \\ 
WASP-46b   & 1.310 & 2.101 & 125.4 \\ 
Kepler-5b   & 1.431 & 2.114 & 106.3 \\ 
HAT-P-22b   & 1.080 & 2.147 & 193.1 \\ 
HAT-P-31b   & 1.070 & 2.171 & 198.6 \\ 
HAT-P-14b   & 1.200 & 2.200 & 153.7 \\ 
KOI-428b   & 1.170 & 2.200 & 162.5 \\ 
WASP-8b   & 1.038 & 2.244 & 217.3 \\ 
CoRoT-21b   & 1.300 & 2.260 & 132.4 \\ 
WASP-36b   & 1.269 & 2.279 & 139.5 \\ 
WASP-66b   & 1.390 & 2.320 & 118.0 \\ 
CoRoT-17b   & 1.020 & 2.450 & 236.9 \\ 
Kepler-17b   & 1.312 & 2.450 & 135.8 \\ 
Qatar-2b   & 1.144 & 2.487 & 182.0 \\ 
CoRoT-11b   & 1.390 & 2.490 & 122.7 \\ 
WASP-53b   & 1.200 & 2.500 & 164.3 \\ 
WASP-38b   & 1.079 & 2.712 & 217.4 \\ 
CoRoT-10b   & 0.970 & 2.750 & 285.3 \\ 
CoRoT-23b   & 1.050 & 2.800 & 235.6 \\ 
CoRoT-6b   & 1.166 & 2.960 & 190.6 \\ 
WASP-10b   & 1.080 & 3.060 & 230.1 \\ 
HD-17156b   & 1.095 & 3.191 & 227.5 \\ 
KOI-135b   & 1.200 & 3.230 & 187.6 \\ 
CoRoT-2b   & 1.465 & 3.310 & 129.9 \\ 
HAT-P-34b   & 1.107 & 3.328 & 226.6 \\ 
CoRoT-18b   & 1.310 & 3.470 & 163.9 \\ 
WASP-32b   & 1.180 & 3.600 & 205.3 \\ 
HD-80606b   & 0.921 & 3.940 & 380.3 \\ 
HAT-P-21b   & 1.024 & 4.063 & 297.8 \\ 
HAT-P-16b   & 1.289 & 4.193 & 186.6
\enddata
\tablecomments{The helium mass fraction has been fixed at 0.25 in the entire envelope. No core has been added. Estimated radii and masses have been taken from \emph{http://www.exoplanet.eu}}
\end{deluxetable}

\clearpage

\begin{deluxetable}{cccc}
\tablewidth{0pt}
\tablenum{6C}
\tablecaption{Characteristic frequency $\nu_0$ (in $\mu Hz$) for a set of coreless models using the estimated radius and mass (here in Jupiter units) of detected giant exoplanets}
\tablehead{
\colhead{Planet}  & \colhead{$R_{planet}$ ($R_J$)}      &
\colhead{$M_{planet}$ ($M_J$)} 
&   \colhead{ $\nu _0$ ($\mu Hz$)} 
}
\startdata
OGLE2-TR-L9b   & 1.614 & 4.340 & 126.5 \\ 
WASP-33b   & 1.438 & 4.590 & 160.1 \\ 
HR-8799b   & 1.100 & 7.000 & 328.2 \\ 
WASP-14b   & 1.281 & 7.341 & 252.4 \\ 
CoRoT-14b   & 1.090 & 7.600 & 347.5 \\ 
KOI-13b   & 1.830 & 8.300 & 158.2 \\ 
Kepler-14b   & 1.136 & 8.400 & 336.8 \\
HAT-P-2b   & 0.951 & 8.740 & 490.3 \\ 
Kepler-30c   & 1.290 & 9.100 & 278.3 \\ 
SWEEPS-11   & 1.130 & 9.700 & 364.8 \\ 
HR-8799c   & 1.300 & 10.000 & 288.1 \\ 
HR-8799d   & 1.200 & 10.000 & 332.0 \\ 
WASP-18b   & 1.165 & 10.430 & 357.5 \\ 
1RXS1609b   & 1.700 & 14.000 & 216.3 \\ 
HN-Pegb   & 1.100 & 16.000 & 487.6 \\ 
Kepler-30d   & 0.960 & 17.000 & 639.1 \\ 
KOI-423b   & 1.220 & 18.000 & 432.3 \\ 
2M-2140+16b   & 0.920 & 20.000 & 742.3 \\ 
GQ-Lupb   & 1.800 & 21.500 & 264.3 \\ 
CoRoT-3b   & 1.010 & 21.660 & 653.3 \\ 
2M-2206-20b   & 1.300 & 30.000 & 501.3 \\ 
2M-0746$+$20b   & 0.970 & 30.000 & 815.6
\enddata
\tablecomments{The helium mass fraction has been fixed at 0.25 in the entire envelope. No core has been added. Estimated radii and masses have been taken from \emph{http://www.exoplanet.eu}}
\end{deluxetable}

\clearpage

\begin{deluxetable}{ccccc}
\tablewidth{0pt}
\tablecaption{Parameters of  coreless exoplanet models of various masses with a radius fixed at $R_P = 1.0$ $R_J$. The helium mass fraction in the envelope is 0.25. \label{parametersMass}}
\tablehead{
\colhead{$M_P$ ($M_J$)}         &
\colhead{$p_c$ ($Mbar$)} & \colhead{$\rho _c$ ($g \ cm^{-3}$)}
&  \colhead{Specific entropy $S$ $(k_B  /  baryon)$}
&   \colhead{ $\nu _0$ ($\mu Hz$)}}
\startdata
0.5   &      9.8     &    2.0    &     6.9   &      107.3\\
1.0   &     39.2    &     4.1    &     6.7    &     157.4\\
 2.0     &    162.5    &     8.6     &    6.5     &    225.8 \\ 
 3.0     &    383.8    &     13.6     &    6.6     &    274.3 \\
 5.0     &    1176    &     24.7     &    7.1     &    345.4 \\ 
10.0     &    5612    &     55.8     &    8.2     &    467.2 \\ 
15.0     &    14075    &     89.1     &    8.9     &    560.5
\enddata
\end{deluxetable}

\begin{deluxetable}{ccccc}
\tablewidth{0pt}
\tablecaption{Parameters of coreless exoplanet models for various radii with a mass fixed ar $M_P = 1.0$ $M_J$. The helium mass fraction in the envelope is 0.25. \label{parametersRadius}}
\tablehead{
\colhead{$R_P$ ($R_J$)}         &
\colhead{$p_c$ ($Mbar$)} & \colhead{$\rho _c$ ($g \ cm^{-3}$)}
&  \colhead{Specific entropy $S$ $(k_B  / baryon)$}
&   \colhead{ $\nu _0$ ($\mu Hz$)}}
\startdata
 0.9     &    48.2    &     5.0     &    3.5     &    220.2 \\ 
1.0   &     39.2    &     4.1    &     6.7    &     157.4\\
1.2     &    28.7    &     3.0     &    8.7     &    102.0\\ 
1.4     &    22.5    &     2.3     &    9.7     &    74.57 \\ 
1.6     &    17.9    &     1.8     &    10.3     &    58.47 \\
1.8     &    14.2    &     1.5     &    10.7     &    47.78 \\ 
2.0     &    11.5    &     1.2     &    10.9     &    40.08
\enddata
\end{deluxetable}

\begin{deluxetable}{ccccc}
\tablewidth{0pt}
\tablecaption{Parameters of exoplanet  models for various core masses $M_c$ with a planet mass fixed at $M_P = 1.0$ $M_J$ and a planet radius fixed at $R_P = 1.0$ $R_J$. The helium mass fraction in the envelope is 0.25. \label{parametersCore1}}
\tablehead{
\colhead{$M_c$ ($M_{\oplus}$)}         &
\colhead{$p_c$ ($Mbar$)} & \colhead{$\rho _c$ ($g \ cm^{-3}$)}
&  \colhead{Specific entropy $S$ $(k_B  / baryon)$}
&   \colhead{ $\nu _0$ ($\mu Hz$)}}
\startdata
0   &     39.2    &     4.1    &     6.7    &     157.4\\
10    &    96.9    &     21.8     &    7.0     &    151.2 \\ 
20    &    145.7    &     25.8     &    7.3     &    149.4 \\ 
30    &    198.9    &     29.2     &    7.6     &    147.5 \\ 
50    &    307.9    &     35.5     &    8.1     &    143.0 \\
100     &    645.4    &     49.5     &    9.2     &    134.1
\enddata
\end{deluxetable}

\begin{deluxetable}{ccccc}
\tablewidth{0pt}
\tablecaption{Parameters of exoplanet  models for various core masses $M_c$ with a planet mass fixed at $M_P = 1.0$ $M_J$ and a specific entropy fixed to 6.67 $k_B  / baryon$ (value for a coreless model with  $R_P = 1.0$ $R_J$ and $M_P = 1.0$ $M_J$). The helium mass fraction in the envelope is 0.25. \label{parametersCore2}}
\tablehead{
\colhead{$M_c$ ($M_{\oplus}$)}    & \colhead{$R_P$ ($R_J$)}
& \colhead{$p_c$ ($Mbar$)} & \colhead{$\rho _c$ ($g \ cm^{-3}$)}
&   \colhead{ $\nu _0$ ($\mu Hz$)}
}
\startdata
0   &  1.0  &   39.2    &     4.1    &  157.4 \\
10    &    0.98    &    100.1    &     23.0     &    160.0 \\ 
20    &  0.96  &    152.2    &     27.3     &    166.8 \\ 
30    &  0.94  &    209.1    &     31.2     &    173.0 \\  
50    &  0.90 &    329.5    &     38.1     &    187.3 \\ 
100    & 0.81 &  680.6    &     53.0      &    223.6
\enddata
\end{deluxetable}

\begin{deluxetable}{ccccc}
\tablewidth{0pt}
\tablecaption{Parameters of coreless exoplanet models for some helium mass fraction $Y$ in the envelope with a planet mass fixed at $M_P = 1.0$ $M_J$ and a planet radius fixed at $R_P = 1.0$ $R_J$. \label{parametersHe}}
\tablehead{
\colhead{$Y$}         &
\colhead{$p_c$ ($Mbar$)} & \colhead{$\rho _c$ ($g \ cm^{-3}$)}
&  \colhead{Specific entropy $S$ $(k_B  / baryon)$}
&   \colhead{ $\nu _0$ ($\mu Hz$)}}
\startdata
0.25   &     39.2    &     4.1    &     6.7    &     157.4\\
0.30    &     41.7    &     4.3     &  6.9     &    153.12
\enddata
\end{deluxetable}

\clearpage

%%%%%%%%%%%%%%%%%%%%%%%%%%%%%%%%%%%%%%%%%%%%%%%%%%%%%%%%%%%%%%%%%%%%%%%%%%%%%%%%%%%%%%%%%%%%%%%

%%%%%%%%%%%%%%%%%%%%%%%%%%%%%

%%%%%%%%%%%%%%%%%%%%%%%%%%%%%

\clearpage

%%%%%%%%%%%%%%%%%%%%%%%%%%%%%%%%%

%%%%%%%%%%%%%%%%%%%%%%%%%%%

\clearpage

\clearpage

%%%%%%%%%%%%%%%%%%%%%%%%%%%

%%%%%%%%%%%%%%%%%%%%%%%%%%%%

\begin{figure}
\plotone{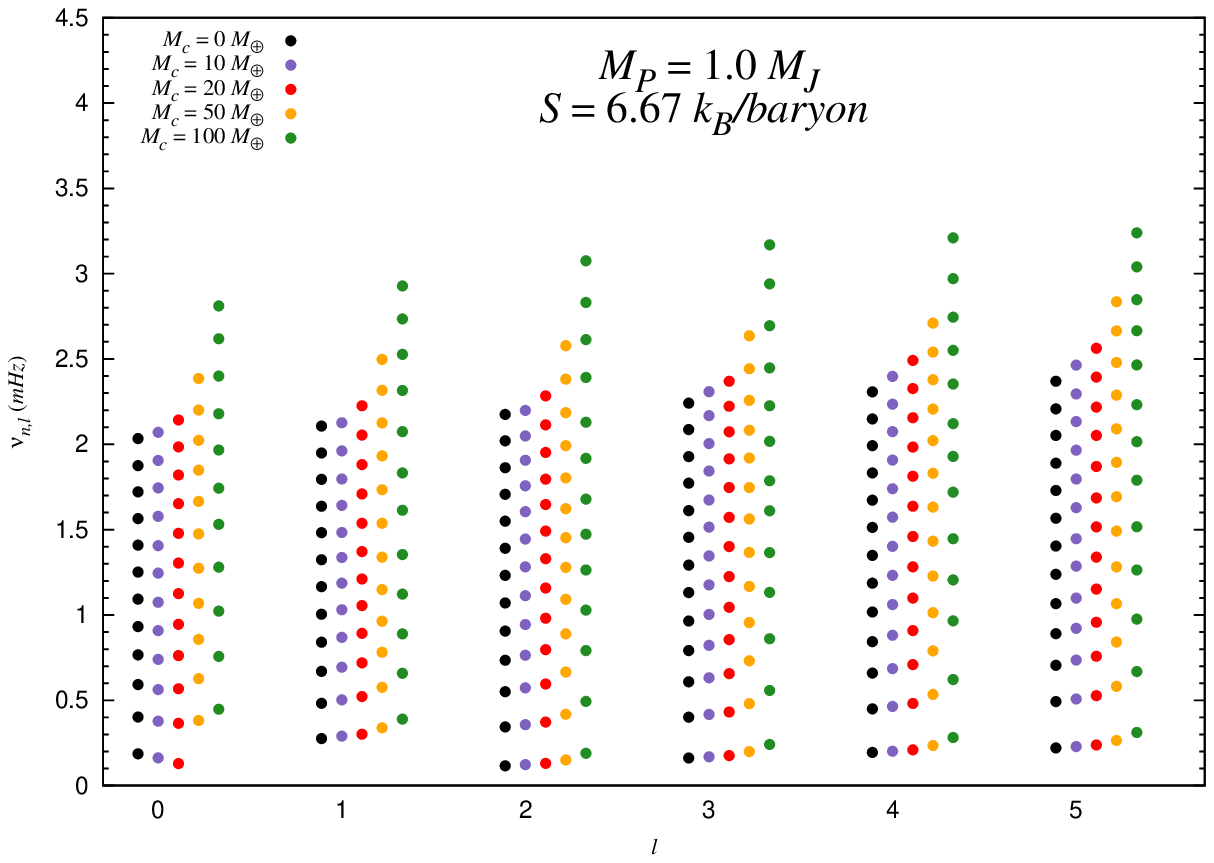}
\caption{Low-order eigenfrequencies of oscillation modes (in $mHz$), as a function of the degree $l$, for exoplanet models with various core masses, with the radius fixed at $R_P = 1.0$ $R_J$ and the specific entropy fixed at $S = 6.67 k_B/baryon$. The helium mass fraction in the envelope is 0.25. For the sake of clarity, the eigenvalue dots of the different models has been seperated and drawn on both sides of each integer degree $l$.  \label{f13} }
\end{figure}

%%%%%%%%%%%%%%%%%%%%%%%%%%%%%%%%%%%%%%%%%%%%%%%%%%%%%%%%%%%%%%%%%%%%%%%%%%%%%%%%%%%%%%%%%%%%%%%

\begin{figure}
\plotone{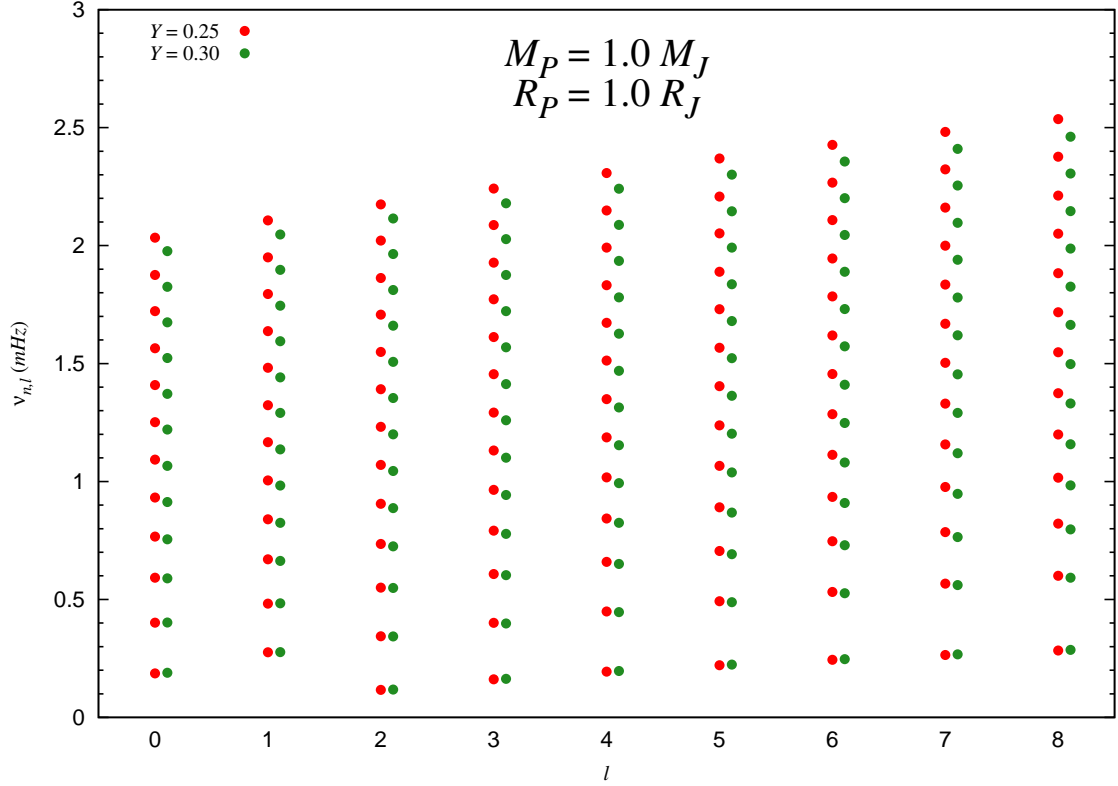}
\caption{Low-order eigenfrequencies of oscillation modes (in $mHz$), as a function of the degree $l$, for exoplanet models with two different helium mass fractions $Y$ in the envelope, with the radius fixed at $R_P = 1.0$ $R_J$ and the mass fixed at $M_P = 1.0$ $M_J$. For the sake of clarity, the eigenvalue dots of the different models has been seperated and drawn on both sides of each integer degree $l$.  \label{f14} }
\end{figure}

%%%%%%%%%%%%%%%%%%%%%%%%%%%%%%%%%%%%%%%%%%%%%%%%%%%%%%%%%%%%%%%%%%%%%%%%%%%%%%%%%%%%%%%%%%%%%%%%%%

\begin{figure}
\includegraphics[width = 0.5\textwidth]{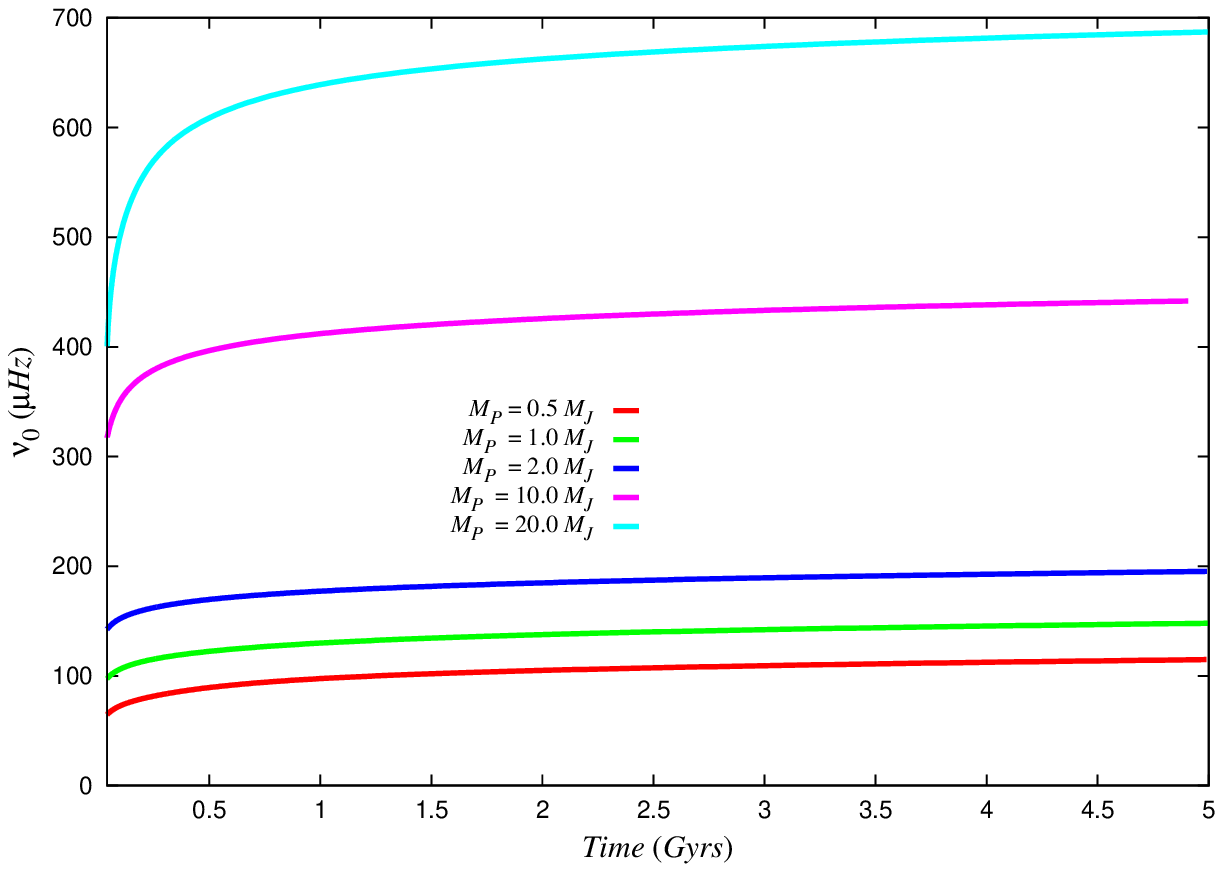}
\includegraphics[width = 0.5\textwidth]{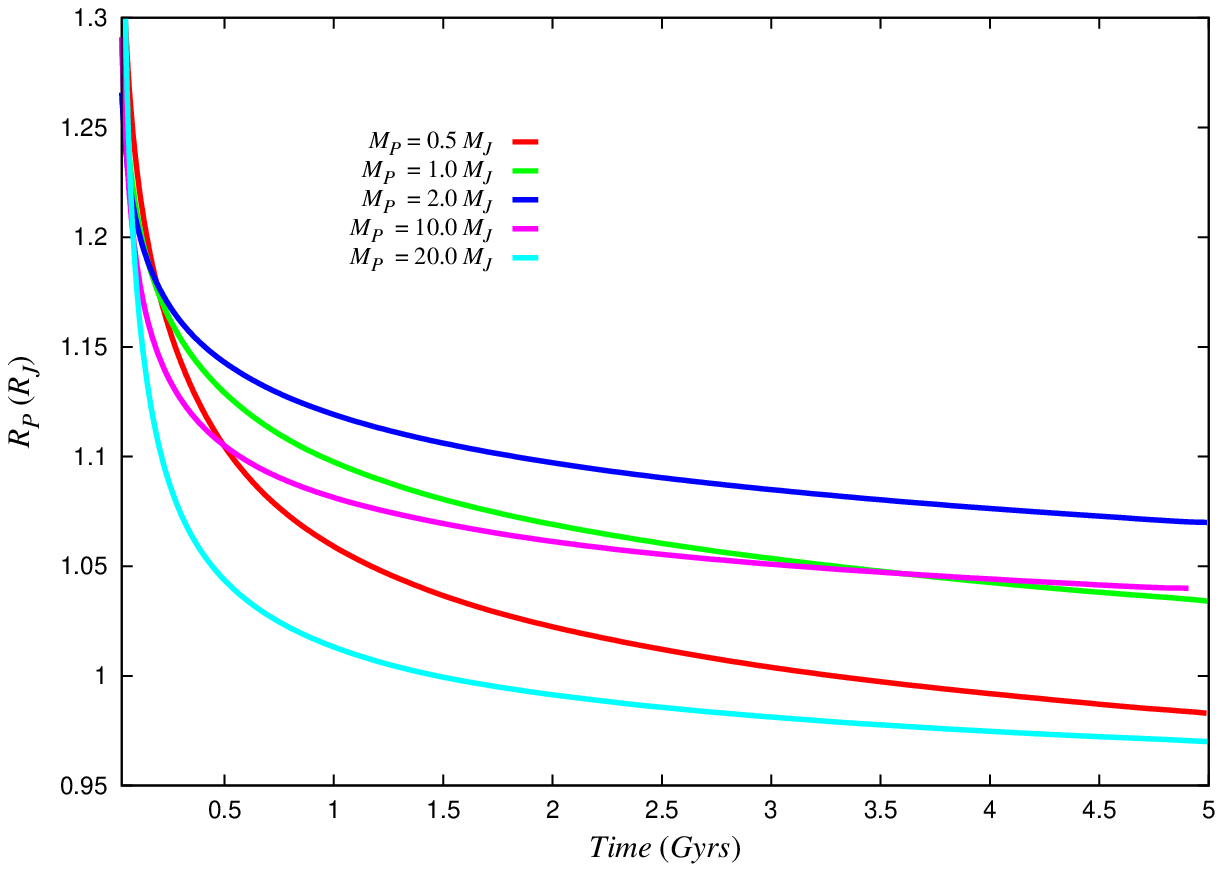}
\caption{Left panel : the charateristic frequency $\nu_0$ as a function of time for various coreless planetary models, characterized by their constant mass ($M_P$ = 0.5, 1.0, 2.0, 10.0 and 20.0$M_J$). The helium mass fraction has been set to 0.25 in the envelope. The planets are considered in isolation during their evolution, which means that no irradiation is taken into account.
Right panel: The corresponding evolution of the planet radius for the same coreless planetary models, in Jupiter units.
The X-axis does not begin at the origin.
\label{f15}
}
\end{figure}

\end{document}